\documentclass[prd,aps,showpacs,epsf,floats,onecolumn]{revtex4}%
\usepackage{amssymb}
\usepackage{amsfonts}
\usepackage{amsmath}
\usepackage{graphicx}%
\providecommand{\U}[1]{\protect\rule{.1in}{.1in}}
\begin{document}
\title{\textbf{Deviations from Geodesic Evolutions and Energy Waste on the Bloch
Sphere}}
\author{\textbf{Leonardo Rossetti}$^{1,2}$, \textbf{Carlo Cafaro}$^{2,3}$,
\textbf{Paul M. Alsing}$^{2}$}
\affiliation{$^{1}$University of Camerino, I-62032 Camerino, Italy}
\affiliation{$^{2}$University at Albany-SUNY, Albany, NY 12222, USA}
\affiliation{$^{3}$SUNY Polytechnic Institute, Utica, NY 13502, USA}

\begin{abstract}
In optimal quantum-mechanical evolutions, motion can occur along
non-predetermined paths of shortest length in an optimal time. Alternatively,
optimal evolutions can happen along predefined paths with no waste of energy
resources and $100\%$ speed efficiency. Unfortunately, realistic physical
scenarios typically result in less-than-ideal evolutions. In this paper, we
study different families of sub-optimal qubit Hamiltonians, both stationary
and time-varying, for which the so-called geodesic efficiency and the speed
efficiency of the corresponding quantum evolutions are less than one.
Furthermore, after proposing an alternative hybrid efficiency measure
constructed out of the two previously mentioned efficiency quantifiers, we
provide illustrative examples where the average departures from
time-optimality and $100\%$ speed efficiency are globally captured over a
limited time period. In particular, thanks to this hybrid measure, quantum
evolutions are partitioned in four categories: Geodesic unwasteful,
nongeodesic unwasteful, geodesic wasteful and, lastly, nongeodesic wasteful.
Finally, we discuss Hamiltonians specified by magnetic field configurations,
both stationary and nonstationary, yielding optimal hybrid efficiency (that
it, both time-optimality and $100\%$ speed efficiency) over a finite time interval.

\end{abstract}

\maketitle

\section{Introduction}

\emph{Background}. It is known that geometric reasoning can be very insightful
in quantum physics \cite{feynman57}. For instance, the geometry of quantum
states can be exploited to specify our restricted capacity in discriminating
one quantum state, either pure or mixed, from another by means of experimental
observations
\cite{provost80,wootters81,mukunda93,braunstein94,braunstein95,karol,crell09}.
In practical scenarios specified by departures from ideal conditions, the
effective quantum-mechanical evolution of a physical system may not be
described by the geometry on the space of quantum states.

First, practical scenarios may not always be specified by evolutions that
occur along the shortest paths, as realistic quantum dynamics evolution could
potentially follow longer paths of evolution due to the presence of possible
imperfections. For instance, not all Hamiltonian evolutions between an initial
and a final pure quantum state yield shortest time evolutions that occur with
optimal speed along the shortest path connecting\textbf{ }a source and a
target state
\cite{anandan90,brody03,carlini06,brody06,brody07,bender07,bender09,ali09,cafaro23,barnes15,barnes21A,barnes21B}%
. For this reason, the actual paths followed by points in projective Hilbert
spaces during quantum dynamics are not the same as the geodesic paths on the
underlying quantum state space with an appropriate metric (e.g., for pure
states, the projective Hilbert space with the Fubini-Study metric). When
focusing on pure states, we can identify Hamiltonian operators that drive
quantum evolutions at the fastest possible rate \cite{bender09,ali09}. These
quantum dynamical trajectories, known as Hamiltonian curves, follow geodesic
paths on the metricized manifolds of quantum states, as states evolve
according to specific physical evolutions. A main quantity used to quantify
the departure from the ideal geodesic evolution on a manifold of pure states
is the so-called geodesic efficiency $\eta_{\mathrm{GE}}$ proposed by Anandan
and Aharonov in Ref. \cite{anandan90}. This quantity is a global quantifier
since it is evaluated over a finite temporal interval. It is defined in terms
of the ratio between two lengths, $s_{0}$ and $s$. The quantity $s_{0}$
denotes the geodesic distance, that is, the length of the shortest path
between the initial and final states ($\left\vert A\right\rangle $ and
$\left\vert B\right\rangle $, respectively) measured with respect to the
Fubini-Study metric (where $s_{0}=2s_{\mathrm{FS}}$, with $s_{\mathrm{FS}}$
being the Fubini-Study distance). The quantity $s$, instead, denotes the
length of the actual path followed by the quantum system. In particular, $s$
depends on the energy uncertainty of the system governed by a stationary or
time-varying Hamiltonian.

Second, practical scenarios may not always be energy-efficient and could
result in some energy wastage. Indeed, instead of minimizing the length of a
non pre-defined path connecting two states, one may be interested in
minimizing the waste of energy resources directly used for the motion of the
quantum system on a pre-defined path in complex projective Hilbert space
\cite{uzdin12,uzdin15,campaioli19,xu24}. In this respect, a key quantity
employed to characterize the effective amount of energy used by the system for
evolving with $100\%$ evolution speed is the so-called speed efficiency
introduced by Uzdin and collaborators in Ref. \cite{uzdin12}. This quantity is
a local quantifier since it is defined at each instant in time. It can be
expressed as the ratio between two energies, the energy uncertainty $\Delta E$
of the system (i.e., the square root of the variance of the generally
time-varying Hamiltonian operator) and the spectral norm $\left\Vert
\mathrm{H}\right\Vert _{\mathrm{SP}}$ of the Hamiltonian operator that
specifies the dynamics.

\medskip

\emph{Physical Motivation}. Considering the information we have presented up
to this point, it is clear that the transition of a quantum system from an
initial source state to a final target state in the minimal time possible,
while simultaneously minimizing energy expenditure, is critically significant
in the realm of quantum information processing. Typically, the time-optimality
of quantum evolution is assessed through the notion of geodesic efficiency,
whereas the energy resources expended during this evolution are evaluated
using the concept of speed efficiency. Additionally, time-optimality can be
defined by examining the progression of the quantum system along an
non-predetermined trajectory in projective Hilbert space over a finite
duration, utilizing a \textquotedblleft global\textquotedblright\ measure such
as geodesic efficiency. Conversely, the least amount of energy expended can be
assessed in real-time via a \textquotedblleft local\textquotedblright\ measure
like the speed efficiency while observing the quantum system as it progresses
along a specified path in projective space. To ensure clarity in the
discussion, consider a two-level quantum system as a spin-$1/2$\textbf{
}particle subjected to an external magnetic field. Then, it is essential to
recognize that optimal ideal conditions for achieving unit geodesic and speed
efficiencies necessitate particular configurations of the magnetic field
corresponding to a specified initial source state on the Bloch sphere.
Furthermore, it is imperative that the magnetic field remains free from any imperfections.

Physics is inherently an experimental discipline characterized by the
prevalence of approximations and imperfections. Specifically, minor
discrepancies in magnetic fields, both in terms of strength and orientation,
are often unavoidable in laboratory settings \cite{tahar20}. For example,
researchers utilize magnetic fields produced by superconducting magnets within
a cyclotron to direct charged particles along intricate trajectories while
maintaining their velocity, as well as to identify these particles by
observing their deflection in the presence of the magnetic field
\cite{tahar20,debnath20,xu20,sharma21}. However, any misalignment of the
magnets may result in alterations to the magnetic field profile. Such
alterations can subsequently lead to instabilities in the specified particle
trajectories, potentially causing deviations from the shortest paths along
with loss of speed.

Driven by these physical considerations, our objective is to leverage geodesic
and speed efficiencies to introduce a concept of hybrid efficiency for quantum
evolutions on the Bloch sphere. This concept aims to quantify both the
deviations from geodesic trajectories and the expenditure of energy resources
over a specified duration.

\medskip

\emph{Goals and Relevance}. From an intuitive standpoint, it is clear that
geodesic efficiency and speed efficiency capture distinct aspects of a quantum
evolution. In particular, it seems reasonable to expect that one would be
naturally interested in achieving both time-optimality (at a global level) and
$100\%$ speed efficiency (at a local level). However, to understand how one
can achieve these two ideal goals, it is important to tackle two main points.
First, one needs to construct suitable families of optimal and suboptimal
quantum evolutions that connect the same quantum states.\ Second, one has to
gain helpful physical insights on the nature of the distinct aspects captured
by these two different optimality measures for quantum evolutions. Therefore,
it is the subject of this paper that of tackling these two relevant points.
Our goal is to address an array of queries, such as:

\begin{enumerate}
\item[{[i]}] How does a time-dependent magnetic field configuration
(specifically, parallel and transverse magnetic field components with respect
to the evolving Bloch vector) change the geodesic and speed efficiencies of a
quantum evolution?

\item[{[ii]}] Can we quantify quantum evolutions that occur over a finite time
interval in terms of both time-optimality and minimum waste of energy
resources by means of a sort of hybrid efficiency measure? How can we
conveniently define such a measure?

\item[{[iii]}] Can we identify suitable time-varying magnetic field
configurations capable of yielding an optimal hybrid efficiency for a quantum evolution?
\end{enumerate}

The discussion of points [i], [ii], and [iii] holds significant importance in
the realm of quantum information and computation for various reasons.
Specifically, a thorough quantitative comprehension of these aspects can
facilitate the creation of effective quantum control strategies that enable
the efficient transfer of a source state to a target state, achieving this in
the least amount of time, at optimal velocity, and with minimal energy
expenditure. Moreover, we anticipate that our hybrid efficiency measure will
prove particularly beneficial in realistic situations where a decision must be
made between optimizing time efficiency or minimizing energy loss, depending
on the acceptable levels of time or energy loss that can be tolerated in
actual experimental settings. Finally, from a theoretical standpoint, we have
reason to believe that this measure may enhance our understanding of the
curvature \cite{carlofuture1} and complexity \cite{carlofuture} of quantum
evolutions. For clarity, lastly, we would like to emphasize that the research
detailed in Ref. \cite{carlofuture1} and the study presented in this paper are
complementary yet separate endeavors, as they address different subjects.

Prior to outlining the structure of this paper, it is important to emphasize
that the terms \textquotedblleft time-dependent, time-varying, and
nonstationary\textquotedblright\ Hamiltonian evolutions (or, conversely,
magnetic field configurations) are utilized interchangeably throughout this study.

The layout of the rest of the paper is as follows. In Section II, after
reviewing the concepts of geodesic and speed efficiencies, we propose a hybrid
efficiency measure $\eta_{\mathrm{HE}}$ suitable for capturing the average
departures from time-optimality and $100\%$ speed efficiency. In Section III,
we construct a one-parameter family of stationary qubit Hamiltonians
connecting two states $\left\vert A\right\rangle $ and $\left\vert
B\right\rangle $ along nongeodesic paths on the Bloch sphere. Furthermore, we
analyze both non-traceless and traceless time-varying Hamiltonians yielding
energy-wasteful evolutions. In both cases, we characterize the evolutions in
terms of the geodesic and speed efficiencies. In Section IV, we use the
formalism introduced in the previous section to present four illustrative
examples. Specifically, we describe the full spectrum of possible quantum
evolutions: Geodesic unwasteful, geodesic wasteful, nongeodesic wasteful and,
finally, nongeodesic unwasteful evolutions. In all cases, we quantify the
average properties of the quantum motion in terms of the newly proposed hybrid
efficiency measure. In Section V, we present a summary of results together
with some final remarks. Finally, technical details are placed in Appendixes A
and B.

\section{Efficiency measures}

In this section, after recalling the concepts of geodesic
\cite{anandan90,cafaro20} and speed \cite{uzdin12} efficiencies, we propose a
hybrid efficiency measure suitable for capturing the average departures from
time-optimality and $100\%$ speed efficiency. In what follows, we limit our
discussion to the physics of two-level quantum systems described by pure
states that evolve under arbitrary Hermitian evolutions.

\subsection{Geodesic efficiency}

We begin by recalling the notion of geodesic efficiency for a quantum
evolution as originally introduced by Anandan and Aharonov in Ref.
\cite{anandan90}. Take into consideration an evolution of a state vector
$\left\vert \psi\left(  t\right)  \right\rangle $ characterized \ by the
time-dependent Schr\"{o}dinger equation, $i\hslash\partial_{t}\left\vert
\psi\left(  t\right)  \right\rangle =\mathrm{H}\left(  t\right)  \left\vert
\psi\left(  t\right)  \right\rangle $, with $t_{A}\leq t\leq t_{B}$. While
this introductory presentation presupposes that the reduced Planck constant,
denoted as $\hslash$, is not equal to one, we will adopt the convention of
setting $\hslash=1$ for the majority of our explicit calculations throughout
this paper. We will make it a point to remind the reader of this choice
whenever it is deemed necessary. Then, the geodesic efficiency $\eta
_{\mathrm{GE}}$ for such a quantum evolution is a time-independent (global)
scalar quantity with $0\leq\eta_{\mathrm{GE}}\leq1$ defined as
\cite{anandan90,cafaro20}%
\begin{equation}
\eta_{\mathrm{GE}}\overset{\text{def}}{=}\frac{s_{0}}{s}=1-\frac{\Delta s}%
{s}=\frac{2\arccos\left[  \left\vert \left\langle A|B\right\rangle \right\vert
\right]  }{2\int_{t_{A}}^{t_{B}}\frac{\Delta E\left(  t\right)  }{\hslash}%
dt}\text{,} \label{efficiency}%
\end{equation}
with $\Delta s\overset{\text{def}}{=}s-s_{0}$. The quantity $s_{0}$ was
mentioned in the Introduction and denotes the distance along the shortest
geodesic path that joins the initial $\left\vert A\right\rangle
\overset{\text{def}}{=}$ $\left\vert \psi\left(  t_{A}\right)  \right\rangle $
and final $\left\vert B\right\rangle \overset{\text{def}}{=}\left\vert
\psi\left(  t_{B}\right)  \right\rangle $ states on the complex projective
Hilbert space. Moreover, as briefly pointed out in our Introduction, the
quantity $s$ in Eq. (\ref{efficiency}) is the distance along the dynamical
trajectory $\gamma\left(  t\right)  :t\mapsto\left\vert \psi\left(  t\right)
\right\rangle $ corresponding to the evolution of the state vector $\left\vert
\psi\left(  t\right)  \right\rangle $ with $t_{A}\leq t\leq t_{B}$. Obviously,
a geodesic quantum evolution with $\gamma\left(  t\right)  =\gamma
_{\mathrm{geodesic}}\left(  t\right)  $ is defined by the relation
$\eta_{\mathrm{GE}}^{(\gamma_{\mathrm{geodesic})}}=1$. Focusing on the
numerator in\ Eq. (\ref{efficiency}), we note that it defines the angle
between the unit state vectors $\left\vert A\right\rangle $ and $\left\vert
B\right\rangle $ and equals the Wootters distance \cite{wootters81}.
Specifically, setting $\rho_{A}\overset{\text{def}}{=}\left\vert
A\right\rangle \left\langle A\right\vert =(\mathbf{1+}\hat{a}\cdot
\mathbf{\boldsymbol{\sigma}})/2$ and $\rho_{B}\overset{\text{def}%
}{=}\left\vert B\right\rangle \left\langle B\right\vert =(\mathbf{1+}\hat
{b}\cdot\mathbf{\boldsymbol{\sigma}})/2$\textbf{ }with unit vectors\textbf{
}$\hat{a}$ and $\hat{b}$ such that\textbf{ }$\hat{a}\cdot$ $\hat{b}%
=\cos(\theta_{AB})$,\textbf{ }it happens that\textbf{ }$s_{0}=\theta_{AB}$
since $\left\vert \left\langle A|B\right\rangle \right\vert ^{2}%
=\mathrm{tr}\left(  \rho_{A}\rho_{B}\right)  +2\sqrt{\det(\rho_{A})\det
(\rho_{B})}=(1+\hat{a}\cdot\hat{b})/2=\cos^{2}\left(  \theta_{AB}/2\right)  $.
Clearly, $\mathbf{\boldsymbol{\sigma}}\overset{\text{def}}{=}\left(
\sigma_{x}\text{, }\sigma_{y}\text{, }\sigma_{z}\right)  $ is the vector
operator specified by the usual Pauli operators. The denominator in Eq.
(\ref{efficiency}), instead, specifies the integral of the infinitesimal
distance $ds\overset{\text{def}}{=}2\left[  \Delta E\left(  t\right)
/\hslash\right]  dt$ along the evolution curve in ray space \cite{anandan90}.
The quantity $\Delta E\left(  t\right)  \overset{\text{def}}{=}\left[
\left\langle \psi|\mathrm{H}^{2}\left(  t\right)  |\psi\right\rangle
-\left\langle \psi|\mathrm{H}\left(  t\right)  |\psi\right\rangle ^{2}\right]
^{1/2}$ is the energy uncertainty of the system expressed as the square root
of the dispersion of $\mathrm{H}\left(  t\right)  $. Notably, Anandan and
Aharonov showed that the infinitesimal distance $ds\overset{\text{def}%
}{=}2\left[  \Delta E\left(  t\right)  /\hslash\right]  dt$ is connected to
the Fubini-Study infinitesimal distance $ds_{\text{\textrm{FS}}}$ via the
relation \cite{anandan90},%
\begin{equation}
ds_{\text{\textrm{FS}}}^{2}\left(  \left\vert \psi\left(  t\right)
\right\rangle \text{, }\left\vert \psi\left(  t+dt\right)  \right\rangle
\right)  \overset{\text{def}}{=}4\left[  1-\left\vert \left\langle \psi\left(
t\right)  |\psi\left(  t+dt\right)  \right\rangle \right\vert ^{2}\right]
=4\frac{\Delta E^{2}\left(  t\right)  }{\hslash^{2}}dt^{2}+\mathcal{O}\left(
dt^{3}\right)  \text{,} \label{relation}%
\end{equation}
where $\mathcal{O}\left(  dt^{3}\right)  $ specifies an infinitesimal quantity
of order equal to or higher than $dt^{3}$. From the connection between
$ds_{\mathrm{FS}}$ and $ds$, one arrives at the conclusion that $s$ is
proportional to the time integral of $\Delta E$. Moreover, $s$ defines the
distance measured by the Fubini-Study metric along the evolution of the
quantum system in ray space. We stress that, when the actual dynamical curve
is the shortest geodesic path joining $\left\vert A\right\rangle $ and
$\left\vert B\right\rangle $, $\Delta s$ equals zero and the geodesic
efficiency $\eta_{\mathrm{GE}}$ in Eq. (\ref{efficiency}) becomes one.
Clearly, $\pi$ is the shortest possible distance between two orthogonal pure
states in ray space. Before moving to the speed efficiency, we note that if we
set \textrm{H}$\left(  t\right)  \overset{\text{def}}{=}h_{0}\left(  t\right)
\mathbf{1+h}\left(  t\right)  \cdot\mathbf{\boldsymbol{\sigma}}$ and
$\rho\left(  t\right)  \overset{\text{def}}{=}(\mathbf{1+}\hat{a}\left(
t\right)  \cdot\mathbf{\boldsymbol{\sigma})/}2$ with $t_{A}\leq t\leq t_{B}$,
the energy uncertainty $\Delta E\left(  t\right)  \overset{\text{def}}{=}%
\sqrt{\mathrm{tr}\left(  \rho\mathrm{H}^{2}\right)  -\left[  \mathrm{tr}%
\left(  \rho\mathrm{H}\right)  \right]  ^{2}}$ reduces to $\Delta E\left(
t\right)  =\sqrt{\mathbf{h}^{2}-\left[  \hat{a}\left(  t\right)
\cdot\mathbf{h}\right]  ^{2}}$. Finally, the geodesic efficiency in Eq.
(\ref{efficiency}) can be recast as%
\begin{equation}
\eta_{\mathrm{GE}}=\frac{2\arccos\left(  \sqrt{\frac{1+\hat{a}\cdot\hat{b}}%
{2}}\right)  }{\int_{t_{A}}^{t_{B}}\frac{2}{\hslash}\sqrt{\mathbf{h}%
^{2}-\left[  \hat{a}\left(  t\right)  \cdot\mathbf{h}\right]  ^{2}}dt}\text{,}
\label{jap}%
\end{equation}
with $\hat{a}\left(  t_{A}\right)  \overset{\text{def}}{=}\hat{a}$ and
$\hat{a}\left(  t_{B}\right)  =\hat{b}$ in Eq. (\ref{jap}). Interestingly, for
$\mathrm{H}\left(  t\right)  \overset{\text{def}}{=}\mathbf{h}\left(
t\right)  \cdot\mathbf{\boldsymbol{\sigma}}$ and setting
\begin{equation}
\mathbf{h}=\left[  \mathbf{h}\cdot\hat{a}\right]  \hat{a}+\left[
\mathbf{h}-(\mathbf{h}\cdot\hat{a})\hat{a}\right]  =\mathbf{h}_{\shortparallel
}+\mathbf{h}_{\perp}\text{,}%
\end{equation}
where $\hat{a}=\hat{a}\left(  t\right)  $ in the decomposition of $\mathbf{h}%
$, $\eta_{\mathrm{GE}}$ in Eq. (\ref{jap}) reduces to
\begin{equation}
\eta_{\mathrm{GE}}=\frac{\arccos\left(  \sqrt{\frac{1+\hat{a}\cdot\hat{b}}{2}%
}\right)  }{\int_{t_{A}}^{t_{B}}h_{\bot}(t)dt}\text{.} \label{goodyo1}%
\end{equation}
Therefore, from Eq. (\ref{goodyo1}) one can see that $\eta_{\mathrm{GE}}$
depends only on $h_{\bot}(t)$. Intriguingly, keeping $\hslash=1$, we note that
Feynman's geometric evolution equation \cite{feynman57} $d\hat{a}%
/dt=2\mathbf{h}\times\hat{a}$ for the time-dependent unit Bloch vector
$\hat{a}=\hat{a}\left(  t\right)  $ can be viewed as a local formulation of
the Anandan-Aharonov relation $s=\int2\Delta E\left(  t\right)  dt$. Indeed,
from $d\hat{a}/dt=2\mathbf{h}\times\hat{a}$, we get $da^{2}=4h_{\bot}%
^{2}dt^{2}$. Similarly, from $s=\int2\Delta E\left(  t\right)  dt$, we obtain
$ds=2h_{\bot}dt$. Therefore, combining these two differential relations, we
finally get $da=2h_{\bot}dt=ds$.

We can now discuss the notion of speed efficiency.

\subsection{Speed efficiency}

In Ref. \cite{uzdin12}, Uzdin et \textit{al}. proposed appropriate families of
nonstationary Hamiltonians capable of generating predetermined dynamical
trajectories characterized by a minimal waste of energy resources. These
trajectories, although being energetically resourceful, are not generally
geodesic paths of minimum length. Specifically, the constraint of minimal
waste of energetic resources is obtained when no energy is wasted on portions
of the Hamiltonian that do not effectively steer the system. Put differently,
all the available energy described by the spectral norm of the Hamiltonian
$\left\Vert \mathrm{H}\right\Vert _{\mathrm{SP}}$ is transformed into the
speed of evolution of the system $v_{\mathrm{H}}(t)$ $\overset{\text{def}%
}{=}(2/\hslash)\Delta E\left(  t\right)  $ with $\Delta E\left(  t\right)  $
being the energy uncertainty. More explicitly, the so-called Uzdin's speed
efficiency $\eta_{\mathrm{SE}}$ is a time-dependent (local) scalar quantity
with $0\leq\eta_{\mathrm{SE}}\leq1$ given by \cite{uzdin12}%
\begin{equation}
\eta_{\mathrm{SE}}\overset{\text{def}}{=}\frac{\Delta\mathrm{H}_{\rho}%
}{\left\Vert \mathrm{H}\right\Vert _{\mathrm{SP}}}=\frac{\sqrt{\mathrm{tr}%
\left(  \rho\mathrm{H}^{2}\right)  -\left[  \mathrm{tr}\left(  \rho
\mathrm{H}\right)  \right]  ^{2}}}{\max\left[  \sqrt{\mathrm{eig}\left(
\mathrm{H}^{\dagger}\mathrm{H}\right)  }\right]  }\text{.} \label{se1}%
\end{equation}
While $\rho=\rho\left(  t\right)  $ is the density operator that describes the
quantum system at time $t$, the quantity $\left\Vert \mathrm{H}\right\Vert
_{\mathrm{SP}}$ in the denominator of Eq. (\ref{se1}) is defined as
$\left\Vert \mathrm{H}\right\Vert _{\mathrm{SP}}\overset{\text{def}}{=}%
\max\left[  \sqrt{\mathrm{eig}\left(  \mathrm{H}^{\dagger}\mathrm{H}\right)
}\right]  $. It is the so-called spectral norm $\left\Vert \mathrm{H}%
\right\Vert _{\mathrm{SP}}$ of the Hamiltonian operator \textrm{H
}which\textrm{ }is a measure of the size of bounded linear operators. It is
expressed as the square root of the maximum eigenvalue of the operator
$\mathrm{H}^{\dagger}\mathrm{H}$, where $\mathrm{H}^{\dagger}$ denotes the
Hermitian conjugate of $\mathrm{H}$. Focusing on two-level quantum systems and
assuming the nonstationary Hamiltonian in Eq. (\ref{se1}) to be given by
$\mathrm{H}\left(  t\right)  \overset{\text{def}}{=}h_{0}\left(  t\right)
\mathbf{1}+\mathbf{h}\left(  t\right)  \cdot\mathbf{\boldsymbol{\sigma}}$, we
observe that the speed efficiency $\eta_{\mathrm{SE}}$ in Eq. (\ref{se1}) can
be conveniently expressed as%
\begin{equation}
\eta_{\mathrm{SE}}=\eta_{\mathrm{SE}}\left(  t\right)  \overset{\text{def}%
}{=}\frac{\sqrt{\mathbf{h}^{2}-(\hat{a}\cdot\mathbf{h})^{2}}}{\left\vert
h_{0}\right\vert +\sqrt{\mathbf{h}^{2}}}\text{.} \label{se2}%
\end{equation}
While $\hat{a}\mathbf{=}\hat{a}\left(  t\right)  $ in Eq. (\ref{se2}) is the
instantaneous unit Bloch vector that describes the qubit state of the system,
the set $\mathrm{eig}\left(  \mathrm{H}^{\dagger}\mathrm{H}\right)  $ in the
definition of $\left\Vert \mathrm{H}\right\Vert _{\mathrm{SP}}$ equals
\begin{equation}
\mathrm{eig}\left(  \mathrm{H}^{\dagger}\mathrm{H}\right)  =\left\{
\lambda_{\mathrm{H}^{\dagger}\mathrm{H}}^{\left(  +\right)  }%
\overset{\text{def}}{=}(h_{0}+\sqrt{\mathbf{h}^{2}})^{2}\text{, }%
\lambda_{\mathrm{H}^{\dagger}\mathrm{H}}^{\left(  -\right)  }%
\overset{\text{def}}{=}(h_{0}-\sqrt{\mathbf{h}^{2}})^{2}\right\}  \text{.}%
\end{equation}
Notice that since the eigenvalues of $\mathrm{H}\left(  t\right)
\overset{\text{def}}{=}h_{0}\left(  t\right)  \mathbf{1}+\mathbf{h}\left(
t\right)  \cdot\mathbf{\boldsymbol{\sigma}}$ are $E_{\pm}\overset{\text{def}%
}{=}h_{0}\pm\sqrt{\mathbf{h\cdot h}}$, the quantity $h_{0}=(E_{+}+E_{-})/2$
represents the average of the two energy levels. Furthermore, $\sqrt
{\mathbf{h}^{2}}=(E_{+}-E_{-})/2$ is proportional to the energy splitting
$E_{+}-E_{-}$ between the two energy levels $E_{\pm}$ with $E_{+}\geq E_{-}$.
Finally, for a trace zero time-dependent Hamiltonian $\mathrm{H}\left(
t\right)  \overset{\text{def}}{=}\mathbf{h}\left(  t\right)  \cdot
\mathbf{\boldsymbol{\sigma}}$ for which $\hat{a}(t)\cdot\mathbf{h}\left(
t\right)  =0$ for any temporal instant $t$, the speed efficiency
$\eta_{\mathrm{SE}}\left(  t\right)  $ is equal to one. Thus, the quantum
evolution happens with no waste of energy resources. Interestingly, for
$\mathrm{H}\left(  t\right)  \overset{\text{def}}{=}\mathbf{h}\left(
t\right)  \cdot\mathbf{\boldsymbol{\sigma}}$ and setting $\mathbf{h}%
=(\mathbf{h}\cdot\hat{a})\hat{a}+\left[  \mathbf{h}-(\mathbf{h}\cdot\hat
{a})\hat{a}\right]  =\mathbf{h}_{\shortparallel}+\mathbf{h}_{\perp}$ with
$\mathbf{h}_{\shortparallel}\cdot\mathbf{h}_{\perp}=0$, $\eta_{\mathrm{SE}%
}\left(  t\right)  $ in Eq. (\ref{se2}) can be completely expressed in term of
the parallel (i.e., $\mathbf{h}_{\shortparallel}\overset{\text{def}%
}{=}h_{\shortparallel}\hat{h}_{\shortparallel}$) and transverse (i.e.,
$\mathbf{h}_{\perp}\overset{\text{def}}{=}h_{\perp}\hat{h}_{\perp}$)
components of the \textquotedblleft magnetic\textquotedblright\ field vector
$\mathbf{h}$ as%
\begin{equation}
\eta_{\mathrm{SE}}\left(  t\right)  =\frac{h_{\bot}(t)}{\sqrt{h_{\bot}%
^{2}(t)+h_{\shortparallel}^{2}(t)}}\text{.} \label{eqtick}%
\end{equation}
From Eq. (\ref{eqtick}), it is evident that $\eta_{\mathrm{SE}}\left(
t\right)  =1$ if and only if $h_{\shortparallel}(t)=0$ for any $t$. In other
words, if and only if $\hat{a}\left(  t\right)  \cdot\mathbf{h}\left(
t\right)  =0$ for any $t$.

Limiting our attention to qubit systems, we point out for completeness that
Hamiltonians $\mathrm{H}\left(  t\right)  $ in Ref. \cite{uzdin12}, which
give\textbf{ }unit speed efficiency trajectories, are proposed so that they
give rise to the same motion $\pi\left(  \left\vert \psi\left(  t\right)
\right\rangle \right)  $ in the complex projective Hilbert space $%
\mathbb{C}
P^{1}$ (or, alternatively, on the Bloch sphere $S^{2}\cong%
\mathbb{C}
P^{1}$) as $\left\vert \psi\left(  t\right)  \right\rangle $. The projection
operator $\pi$ can be formally defined in terms of $\pi:\mathcal{H}_{2}^{1}%
\ni\left\vert \psi\left(  t\right)  \right\rangle \mapsto\pi\left(  \left\vert
\psi\left(  t\right)  \right\rangle \right)  \in%
\mathbb{C}
P^{1}$. It can be demonstrated that an Hamiltonian $\mathrm{H}\left(
t\right)  $ of this kind can be expressed as \cite{uzdin12}%
\begin{equation}
\mathrm{H}\left(  t\right)  =i\left\vert \partial_{t}m(t)\right\rangle
\left\langle m(t)\right\vert -i\left\vert m(t)\right\rangle \left\langle
\partial_{t}m(t)\right\vert \text{,} \label{optH}%
\end{equation}
where, for ease of notation, we can put $\left\vert m(t)\right\rangle
=\left\vert m\right\rangle $ and $\left\vert \partial_{t}m(t)\right\rangle
=\left\vert \partial_{t}m\right\rangle =\partial_{t}\left\vert m\right\rangle
=\left\vert \dot{m}\right\rangle $. The state $\left\vert m\right\rangle $
satisfies $\pi\left(  \left\vert m(t)\right\rangle \right)  =\pi\left(
\left\vert \psi\left(  t\right)  \right\rangle \right)  $, $i\partial
_{t}\left\vert m(t)\right\rangle =\mathrm{H}\left(  t\right)  \left\vert
m(t)\right\rangle $, and $\eta_{\mathrm{SE}}\left(  t\right)  =1$. The
constraint $\pi\left(  \left\vert m(t)\right\rangle \right)  =\pi\left(
\left\vert \psi\left(  t\right)  \right\rangle \right)  $ entails that
$\left\vert m(t)\right\rangle =c(t)\left\vert \psi\left(  t\right)
\right\rangle $, with $c(t)$ being a complex scalar function. Requiring that
$\left\langle m\left\vert m\right.  \right\rangle =1$, we obtain $\left\vert
c(t)\right\vert =1$. Therefore, $c(t)$ can be recast as $e^{i\phi\left(
t\right)  }$ with the phase $\phi\left(  t\right)  \in%
\mathbb{R}
$. Then, enforcing the parallel transport condition $\left\langle m\left\vert
\dot{m}\right.  \right\rangle =\left\langle \dot{m}\left\vert m\right.
\right\rangle =0$, the phase $\phi\left(  t\right)  $ reduces to
$i\int\left\langle \psi\left\vert \dot{\psi}\right.  \right\rangle dt$.
Therefore, the state $\left\vert m(t)\right\rangle $ becomes $\left\vert
m(t)\right\rangle =\exp(-\int_{0}^{t}\left\langle \psi(t^{\prime})\left\vert
\partial_{t^{\prime}}\psi(t^{\prime})\right.  \right\rangle dt^{\prime
})\left\vert \psi\left(  t\right)  \right\rangle $. Observe that the
Hamiltonian $\mathrm{H}\left(  t\right)  $ in Eq. (\ref{optH}) has trace zero
by construction given that its matrix representation with respect to the in
the orthogonal basis $\left\{  \left\vert m\right\rangle \text{, }\left\vert
\partial_{t}m\right\rangle \right\}  $ has only off-diagonal elements.
Moreover, the equation $i\partial_{t}\left\vert m(t)\right\rangle
=\mathrm{H}\left(  t\right)  \left\vert m(t)\right\rangle $ suggests that
$\left\vert m(t)\right\rangle $ fulfils the time-dependent Schr\"{o}dinger
evolution equation. Lastly, the condition $\eta_{\mathrm{SE}}\left(  t\right)
=1$ means that $\mathrm{H}\left(  t\right)  $ steers the state $\left\vert
m(t)\right\rangle $ with maximal speed and no waste of energetic resources.

Having recalled the notions of geodesic and speed efficiencies, we are now
ready to propose our hybrid efficiency measure.\begin{table}[t]
\centering
\begin{tabular}
[c]{c|c|c|c|c}\hline\hline
\textbf{Type of scenario} & $\bar{\eta}_{\mathrm{GE}}$ & $\bar{\eta
}_{\mathrm{SE}}$ & $\eta_{\mathrm{HE}}$ & \textbf{Type of quantum
evolution}\\\hline
First & $1$ & $1$ & $1$ & Geodesic unwasteful\\\hline
Second & $<1$ & $1$ & $<1$ & Nongeodesic unwasteful\\\hline
Third & $1$ & $<1$ & $<1$ & Geodesic wasteful\\\hline
Fourth, first subcase & $<1$ & $\ll1$ & $\ll1$ & More wasteful than
nongeodesic\\\hline
Fourth, second subcase & $\ll1$ & $<1$ & $\ll1$ & Less wasteful than
nongeodesic\\\hline
Fourth, third subcase & $<1$ & $<1$ & $<1$ & As wasteful as
nongeodesic\\\hline
\end{tabular}
\caption{Schematic summary of the variety of scenarios for quantum evolutions
in terms of the average departure from geodesic motion (quantified by the
average geodesic efficiency $\bar{\eta}_{\mathrm{GE}}$ over $\left[
t_{A}\text{, }t_{B}\right]  $) and of the average deviation from minimally
energy wasteful evolutions (quantified by the average speed efficiency
$\bar{\eta}_{\mathrm{SE}}$ over $\left[  t_{A}\text{, }t_{B}\right]  $).
Finally, the hybrid efficiency $\eta_{\mathrm{HE}}$ is given by $\eta
_{\mathrm{HE}}\protect\overset{\text{def}}{=}\bar{\eta}_{\mathrm{GE}}\cdot
\bar{\eta}_{\mathrm{SE}}$.}%
\end{table}

\subsection{Hybrid efficiency}

In what follows, we focus on single-qubit quantum Hamiltonian evolutions on
the Bloch sphere. We have discussed two distinct efficiency measures of
quantum evolutions. The first measure, called \emph{geodesic efficiency }%
$\eta_{\mathrm{GE}}\overset{\text{def}}{=}s_{0}/s$, specifies the departure of
the actual quantum path of length $s$ connecting two given unit quantum states
$\left\vert A\right\rangle $ and $\left\vert B\right\rangle $ from the ideal
geodesic path of length $s_{0}$. The concept of length of a path, be it
$s_{0}$ calculated using the Fubini-Study metric or $s$ calculated according
to the Anandan-Aharonov prescription, is the key notion in the definition of
$\eta_{\mathrm{GE}}$. Therefore, the shortest path connecting two pre-defined
quantum states is the maximally efficient according to $\eta_{\mathrm{GE}}$.
When using $\eta_{\mathrm{GE}}$, we can only speak of geodesic or nongeodesic
quantum evolutions that occur during a temporal interval of length
$t_{B}-t_{A}$. It does not make sense to speak of the geodesic efficiency at
given instant in time. The efficiency $\eta_{\mathrm{GE}}$ is not an
instantaneous quantity. Rather, it specifies a global property of the quantum
evolution. The second measure, called \emph{speed efficiency }$\eta
_{\mathrm{SE}}\left(  t\right)  \overset{\text{def}}{=}\Delta E\left(
t\right)  /\left\Vert \mathrm{H}(t)\right\Vert _{\mathrm{SP}}$ is an
instantaneous quantity.

The key quantity in its definition is the concept of energy or, alternatively,
the concept of speed of a quantum evolution in projective Hilbert space given
by $v_{\mathrm{H}}(t)\overset{\text{def}}{=}\Delta E\left(  t\right)
/\hslash$ (as defined in Ref. \cite{uzdin12}). Therefore, when the speed of
quantum evolution of the system equals the maximal speed (i.e., $\left\Vert
\mathrm{H}(t)\right\Vert _{\mathrm{SP}}/\hslash$) allotted by the available
energetic resources, a maximally speed efficient quantum evolution occurs
according to the measure $\eta_{\mathrm{SE}}\left(  t\right)  $. When
employing $\eta_{\mathrm{SE}}\left(  t\right)  $, it only makes sense speaking
of energetically unwasteful or energetically wasteful quantum evolutions on a
pre-defined and non necessarily geodesic quantum path on the Bloch sphere.

Given the global quantity $\eta_{\mathrm{GE}}\left(  t_{A}\text{, }%
t_{B}\right)  \overset{\text{def}}{=}2\arccos\left[  \left\vert \left\langle
A\left\vert B\right.  \right\rangle \right\vert \right]  /\int_{t_{A}}^{t_{B}%
}(2/\hslash)\Delta E(t)dt$ and the local quantity $\eta_{\mathrm{SE}}\left(
t\right)  \overset{\text{def}}{=}\Delta E\left(  t\right)  /\left\Vert
\mathrm{H}(t)\right\Vert _{\mathrm{SP}}$ with $t_{A}\leq t\leq t_{B}$, we wish
to define a (global) hybrid measure of efficiency that takes into
consideration both deviations from trajectories of shortest length and
departures from ideal energetically unwasteful quantum dynamical trajectories
on the Bloch sphere. Furthermore, we require this hybrid efficiency measure
denoted as $\eta_{\mathrm{HE}}$ to be such that: i) $0\leq\eta_{\mathrm{HE}%
}\leq1$; ii) $\eta_{\mathrm{HE}}=\eta_{\mathrm{HE}}\left(  \bar{\eta
}_{\mathrm{GE}}\text{, }\bar{\eta}_{\mathrm{SE}}\right)  $; iii)
$\eta_{\mathrm{HE}}\rightarrow\bar{\eta}_{\mathrm{GE}}$ when $\bar{\eta
}_{\mathrm{SE}}\rightarrow1$; iv) $\eta_{\mathrm{HE}}\rightarrow\bar{\eta
}_{\mathrm{SE}}$ when $\bar{\eta}_{\mathrm{GE}}\rightarrow1$; v)
$\eta_{\mathrm{HE}}\leq\min\left\{  \bar{\eta}_{\mathrm{GE}}\text{, }\bar
{\eta}_{\mathrm{SE}}\right\}  $. The quantities $\bar{\eta}_{\mathrm{GE}}$ and
$\bar{\eta}_{\mathrm{SE}}$ are given by%
\begin{equation}
\bar{\eta}_{\mathrm{GE}}=\bar{\eta}_{\mathrm{GE}}\left(  t_{A}\text{, }%
t_{B}\right)  \overset{\text{def}}{=}\frac{1}{t_{B}-t_{A}}\int_{t_{A}}^{t_{B}%
}\eta_{\mathrm{GE}}(t)dt\text{,}%
\end{equation}
with $\eta_{\mathrm{GE}}\left(  t\right)  \overset{\text{def}}{=}%
2\arccos\left[  \left\vert \left\langle A\left\vert \psi\left(  t\right)
\right.  \right\rangle \right\vert \right]  /\int_{t_{A}}^{t}(2/\hslash)\Delta
E(t^{\prime})dt^{\prime}$, and%
\begin{equation}
\bar{\eta}_{\mathrm{SE}}=\bar{\eta}_{\mathrm{SE}}\left(  t_{A}\text{, }%
t_{B}\right)  \overset{\text{def}}{=}\frac{1}{t_{B}-t_{A}}\int_{t_{A}}^{t_{B}%
}\eta_{\mathrm{SE}}(t)dt\text{,}%
\end{equation}
respectively. The quantity $\bar{\eta}_{\mathrm{GE}}\left(  t_{A}\text{,
}t_{B}\right)  $ is the average geodesic efficiency of the quantum evolution
over the time interval $\left[  t_{A}\text{, }t_{B}\right]  $. It quantifies
the average deviation of the actual evolution from the geodesic evolution. The
quantity $\bar{\eta}_{\mathrm{SE}}\left(  t_{A}\text{, }t_{B}\right)  $,
instead, is the average speed efficiency of the quantum evolution over the
time interval $\left[  t_{A}\text{, }t_{B}\right]  $. It describes the average
waste of energy resources during the evolution from $t_{A}$ to $t_{B}$. For
completeness, we remark that when we define the average speed efficiency
$\bar{\eta}_{\mathrm{SE}}$, we are in the presence of a reasonable and global
extension of a truly local quantity defined at an instant $t$ \textbf{(}i.e.,
$\eta_{\mathrm{SE}}\left(  t\right)  $\textbf{). }Unlike $\eta_{\mathrm{SE}%
}\left(  t\right)  $, $\eta_{\mathrm{GE}}\left(  t\right)  $ is a sort of
global quantity since its definition requires the specification of a finite
temporal interval $\left[  t_{A}\text{, }t\right]  $. Therefore, when
evaluating the average geodesic efficiency\textbf{ }$\bar{\eta}_{\mathrm{GE}}%
$,\textbf{ }one needs to take into account the fact that there is an
underlying choice of an initial time $t_{A}$, with\textbf{ }$t_{A}\leq
t^{\prime}\leq t\leq t_{B}$ as clear from the definitions of $\eta
_{\mathrm{GE}}\left(  t\right)  $ and $\bar{\eta}_{\mathrm{GE}}=\bar{\eta
}_{\mathrm{GE}}\left(  t_{A}\text{, }t_{B}\right)  $. The imposition of the
above-mentioned five conditions can be physically explained as follows. To
begin, condition i) simply demands that the numerical values assumed
by\textbf{ }$\eta_{\mathrm{HE}}$\textbf{ }belong to the same range as that
specifying the efficiencies $\eta_{\mathrm{GE}}$\textbf{ }and\textbf{ }%
$\eta_{\mathrm{SE}}$\textbf{ }(or, alternatively, $\bar{\eta}_{\mathrm{GE}}%
$\textbf{ }and\textbf{ }$\bar{\eta}_{\mathrm{SE}}$\textbf{).} Condition ii)
expresses the reasonable fact that the functional form of\textbf{ }%
$\eta_{\mathrm{HE}}$\textbf{ }must be expressible in terms of\textbf{ }%
$\bar{\eta}_{\mathrm{GE}}$\textbf{ }and\textbf{ }$\bar{\eta}_{\mathrm{SE}}%
$\textbf{. }Furthermore, conditions iii) and iv) require that $\eta
_{\mathrm{HE}}$\textbf{ }reduces to\textbf{ }$\bar{\eta}_{\mathrm{GE}}%
$\textbf{ }when there is no average waste of energy or, alternatively,
to\textbf{ }$\bar{\eta}_{\mathrm{SE}}$\textbf{ }in the absence of any average
deviation from a geodesic motion on the Bloch sphere. Finally, since the
hybrid efficiency is built so that encodes information about a quantum
evolution both in terms of nongeodesicity and energy resources, one expects
that it will be generally more demanding (with respect to the
efficiencies\textbf{ }$\bar{\eta}_{\mathrm{GE}}$\textbf{ }and\textbf{ }%
$\bar{\eta}_{\mathrm{SE}}$\textbf{, }each encoding only one aspect of the
quantum evolution) to achieve a high value of $\eta_{\mathrm{HE}}$\textbf{.
}Therefore, we impose\textbf{ }$\eta_{\mathrm{HE}}\leq\min\left\{  \bar{\eta
}_{\mathrm{GE}}\text{, }\bar{\eta}_{\mathrm{SE}}\right\}  $\textbf{.}

Given these five requirements, we propose to consider the following hybrid
efficiency measure
\begin{equation}
\eta_{\mathrm{HE}}\left(  t_{A}\text{, }t_{B}\right)  \overset{\text{def}%
}{=}\bar{\eta}_{\mathrm{GE}}\left(  t_{A}\text{, }t_{B}\right)  \bar{\eta
}_{\mathrm{SE}}\left(  t_{A}\text{, }t_{B}\right)  \label{kelly}%
\end{equation}
By construction, $\eta_{\mathrm{HE}}\left(  t_{A}\text{, }t_{B}\right)  $ in
Eq. (\ref{kelly}) satisfies properties i), ii), iii), iv), and v). Note that
one may think of considering the arithmetic and the geometric means between
$\bar{\eta}_{\mathrm{GE}}$ and $\bar{\eta}_{\mathrm{SE}}$ given by%
\begin{equation}
\frac{\bar{\eta}_{\mathrm{GE}}+\bar{\eta}_{\mathrm{SE}}}{2}\text{, and }%
\sqrt{\bar{\eta}_{\mathrm{GE}}\bar{\eta}_{\mathrm{SE}}}\text{,}%
\end{equation}
respectively. However, both proposals fail to satisfy properties iii), iv),
and v). The newly proposed hybrid efficiency measure $\eta_{\mathrm{HE}%
}\left(  t_{A}\text{, }t_{B}\right)  $ in Eq. (\ref{kelly}) specifies the
average deviation of the actual evolution from the geodesic evolution together
with the average waste of energy resources over the time interval $\left[
t_{A}\text{, }t_{B}\right]  $. More explicitly, its interpretation can be
explained by considering the following scenarios specified in terms of the
original pair of values $\left(  \bar{\eta}_{\mathrm{GE}}\text{, }\bar{\eta
}_{\mathrm{SE}}\right)  $. In the first scenario, assume $\left(  \bar{\eta
}_{\mathrm{GE}}\text{, }\bar{\eta}_{\mathrm{SE}}\right)  =\left(  1\text{,
}1\right)  $. In this case, $\eta_{\mathrm{HE}}=1$ and we deal with geodesic
unwasteful paths. In the second scenario, assume $\left(  \bar{\eta
}_{\mathrm{GE}}\text{, }\bar{\eta}_{\mathrm{SE}}\right)  =\left(  <1\text{,
}1\right)  $. In this case, $\eta_{\mathrm{HE}}=\bar{\eta}_{\mathrm{GE}}<1$
and we deal with nongeodesic unwasteful paths. In the third scenario, assume
$\left(  \bar{\eta}_{\mathrm{GE}}\text{, }\bar{\eta}_{\mathrm{SE}}\right)
=\left(  1\text{, }<1\right)  $. In this case $\eta_{\mathrm{HE}}=\bar{\eta
}_{\mathrm{SE}}<1$, and we deal with geodesic wasteful paths. In the fourth
scenario, assume $\left(  \bar{\eta}_{\mathrm{GE}}\text{, }\bar{\eta
}_{\mathrm{SE}}\right)  =\left(  <1\text{, }<1\right)  $. In this case, we
have three subcases. Before discussing these three subcases, note that
$\bar{\eta}_{\mathrm{GE}}=1-\left\langle \Delta s/s\right\rangle _{\Delta t}$
and $\bar{\eta}_{\mathrm{SE}}=1-\left\langle \Delta\epsilon/\epsilon
\right\rangle _{\Delta t}$, with $\Delta t\overset{\text{def}}{=}t_{B}-t_{A}$,
$\Delta s\left(  t\right)  \overset{\text{def}}{=}s(t)-s_{0}(t)$,
$\Delta\epsilon\left(  t\right)  \overset{\text{def}}{=}\left\Vert
\mathrm{H}(t)\right\Vert _{\mathrm{SP}}-\Delta E\left(  t\right)  $, and
$\epsilon\left(  t\right)  \overset{\text{def}}{=}\left\Vert \mathrm{H}%
(t)\right\Vert _{\mathrm{SP}}$. Then, in the first subcase, assume
$\Delta\epsilon/\epsilon>\Delta s/s$. We deal then with more wasteful than
nongeodesic paths. In the second subcase, assume $\Delta\epsilon
/\epsilon<\Delta s/s$. We deal then with less wasteful than nongeodesic paths.
In the third subcase, finally, assume $\Delta\epsilon/\epsilon=\Delta s/s$. We
then deal with as wasteful as nongeodesic paths. In conclusion, by means of
$\eta_{\mathrm{HE}}\left(  t_{A}\text{, }t_{B}\right)  $, we can characterize
quantum evolutions in terms of average geodesicity, average energetic waste,
and any combination of these two average properties of quantum evolutions over
a finite time interval $\left[  t_{A}\text{, }t_{B}\right]  $. The hybrid
measure $\eta_{\mathrm{HE}}\left(  t_{A}\text{, }t_{B}\right)  $ mimics the
behavior of the lowest efficiency measure between $\bar{\eta}_{\mathrm{GE}}$
and $\bar{\eta}_{\mathrm{SE}}$. Specifically, we have the following. In the
first scenario, $\eta_{\mathrm{HE}}=\bar{\eta}_{\mathrm{GE}}=\bar{\eta
}_{\mathrm{SE}}=1$ and we are dealing with a geodesic unwasteful evolution. In
the second scenario, $\eta_{\mathrm{HE}}=\bar{\eta}_{\mathrm{GE}}<1$ and we
are dealing with a nongeodesic unwasteful evolution. In the third scenario,
$\eta_{\mathrm{HE}}=\bar{\eta}_{\mathrm{SE}}<1$ and we are dealing with a
geodesic wasteful evolution. In the first subcase of the fourth case,
$\bar{\eta}_{\mathrm{GE}}<1$ and $\bar{\eta}_{\mathrm{SE}}\ll1$. Then,
$\eta_{\mathrm{HE}}\ll1$ and we are dealing with more wasteful than
nongeodesic evolutions. In the second subcase of the fourth case, since
$\bar{\eta}_{\mathrm{GE}}\ll1$ and $\bar{\eta}_{\mathrm{SE}}<1$,
$\eta_{\mathrm{HE}}\ll1$ and we are dealing with less wasteful than
nongeodesic evolutions. Finally, in the third subcase of the fourth case, we
have $\bar{\eta}_{\mathrm{GE}}=$ $\bar{\eta}_{\mathrm{SE}}<1$. Then
$\eta_{\mathrm{HE}}\leq\bar{\eta}_{\mathrm{GE}}=\bar{\eta}_{\mathrm{SE}}<1$
and we are dealing with evolutions that are as wasteful as nongeodesic. We
refer to Table I for a visual summary of all the quantum-mechanical scenarios
discussed here.

Having introduced the geodesic, speed, and hybrid efficiencies of quantum
evolutions on a Bloch sphere in Eqs. (\ref{efficiency}), (\ref{se1}), and
(\ref{kelly}), respectively, we can now discuss sub-optimal quantum evolution
scenarios in the next section.

\section{Deviations from ideality}

In this section, we begin by constructing a one-parameter family of stationary
qubit Hamiltonians connecting two states $\left\vert A\right\rangle $ and
$\left\vert B\right\rangle $ along nongeodesic paths on the Bloch sphere. We
then analyze both non-traceless and traceless nonstationary Hamiltonians
yielding energy-wasteful evolutions. In both cases, we quantify aspects of the
quantum evolutions in terms of the geodesic and speed efficiencies introduced
in the previous section.\begin{figure}[t]
\centering
\includegraphics[width=0.5\textwidth] {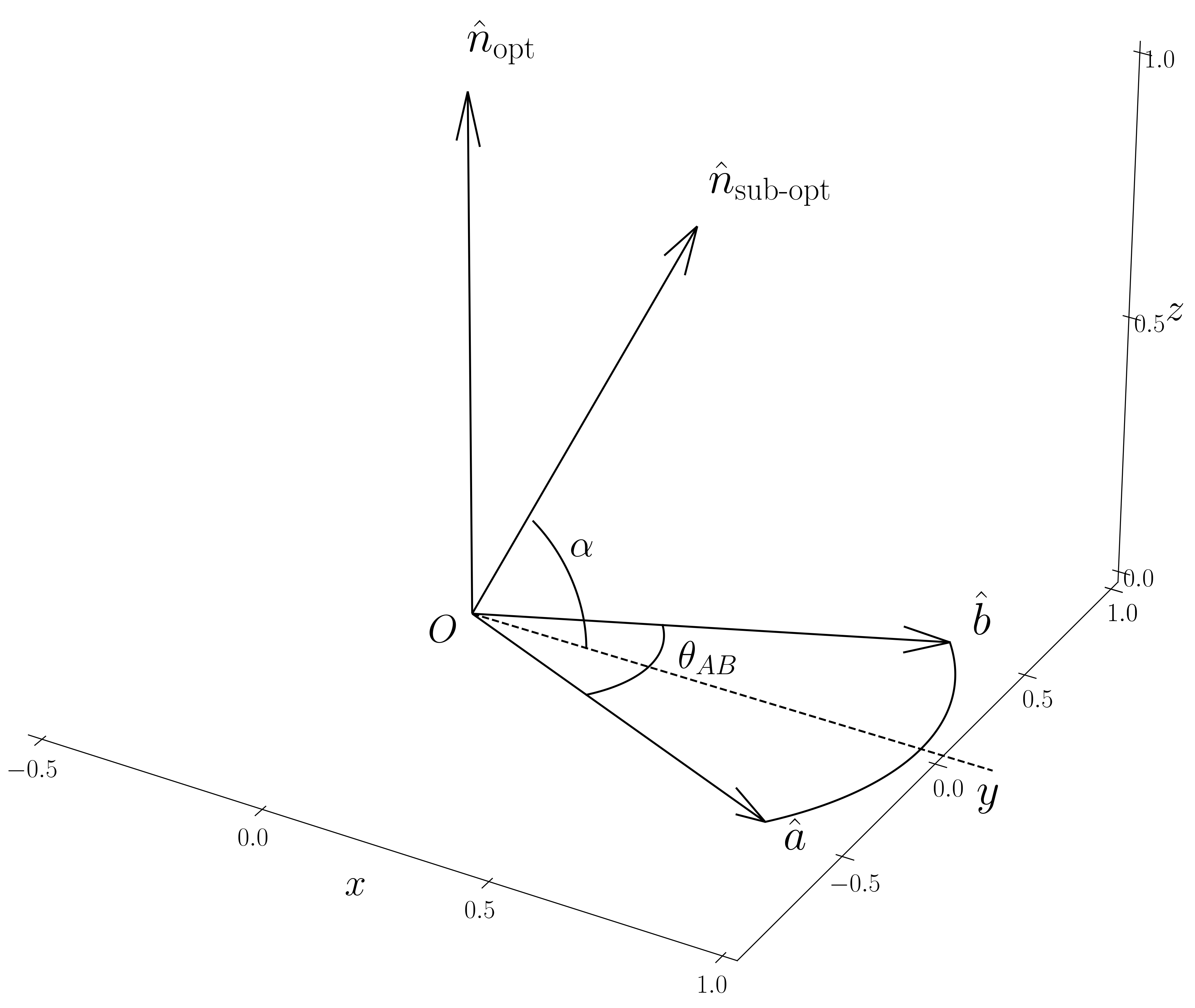}\caption{Schematic depiction of
the unit vectors $\hat{n}_{\mathrm{opt}}\protect\overset{\text{def}}{=}$
$(\hat{a}\times\hat{b})/\left\Vert \hat{a}\times\hat{b}\right\Vert $ and
$\hat{n}_{\mathrm{sub}\text{-}\mathrm{opt}}\protect\overset{\text{def}}{=}%
\cos(\alpha)\left[  (\hat{a}+\hat{b})/\left\Vert \hat{a}+\hat{b}\right\Vert
\right]  +\sin(\alpha)\left[  (\hat{a}\times\hat{b})/\left\Vert \hat{a}%
\times\hat{b}\right\Vert \right]  $ specifying the optimal and sub-optimal
rotation axes, respectively. The initial and final Bloch vectors $\hat{a}$ and
$\hat{b}$, respectively, are such that $\hat{a}\cdot\hat{b}=\cos(\theta_{AB}%
)$.}%
\end{figure}

\subsection{From geodesic to nongeodesic evolutions}

In this subsection, we focus on how to generate sub-optimal unitary operators
$U_{\text{\textrm{sub-opt}}}(t)$ such that $U_{\text{\textrm{sub-opt}}}%
(t_{AB})|A\rangle=|B\rangle$ for arbitrary pure qubit states $|A\rangle$ and
$|B\rangle$ with $t_{AB}$ greater than the minimal evolution time that
specifies optimal quantum evolutions. In particular, we focus on unitary
operators that emerge from stationary traceless Hamiltonians. Essentially,
these unitary time propagators act on qubits as rotations around a fixed axis
in the Bloch sphere. More specifically, $U_{\text{\textrm{sub-opt}}}%
(t_{AB})=e^{-\frac{i}{\hslash}\mathrm{H}_{\text{\textrm{sub-opt}}}t_{AB}%
}=e^{-i\frac{\phi}{2}\hat{n}_{\mathrm{sub}\text{\textrm{-}}\mathrm{opt}}%
\cdot\mathbf{\boldsymbol{\sigma}}}$. Therefore, we need to find a rotation
angle $\phi$ and a rotation axis $\hat{n}_{\mathrm{sub}\text{\textrm{-}%
}\mathrm{opt}}$ with $\mathrm{H}_{\text{\textrm{sub-opt}}}\overset{\text{def}%
}{=}E\hat{n}_{\mathrm{sub}\text{\textrm{-}}\mathrm{opt}}\cdot
\mathbf{\boldsymbol{\sigma}}$, $(E/\hslash)t_{AB}=\phi/2$, and $E$ expressed
in energy units. We point out that, exploiting the correspondence between unit
quantum states evolving on the Bloch sphere and unit vectors evolving on a
three-dimensional unit sphere, all the results that we obtain in what follows
rely on three-dimensional geometry and simple trigonometry. To better explain
our geometric reasoning in the following discussions, we refer to Fig. $1$ and
Fig. $2$.

We remark that the derivations contained in this subsection A can be a bit
tedious in terms of mathematical details, despite being rather interesting for
those intrigued by geometric perspectives on quantum mechanics. The reader,
however, could avoid its reading and proceed to subsection B provided that
Eqs. (\ref{local}) and (\ref{yoyo}) are taken into considerations. Eqs.
(\ref{local}) and (\ref{yoyo}) are the two main results of subsection A, and
are needed to evaluate the geodesic efficiency $\eta_{\mathrm{GE}}$. They
express the travel time $t_{AB}\left(  \alpha\right)  $ and the length of the
path $s\left(  \alpha\right)  $\textbf{, }respectively, that correspond to the
sub-optimal quantum evolution specified by $\mathrm{H}_{\text{\textrm{sub-opt}%
}}$ in Eq. (\ref{SUB}).

\begin{figure}[t]
\centering
\includegraphics[width=0.5\textwidth] {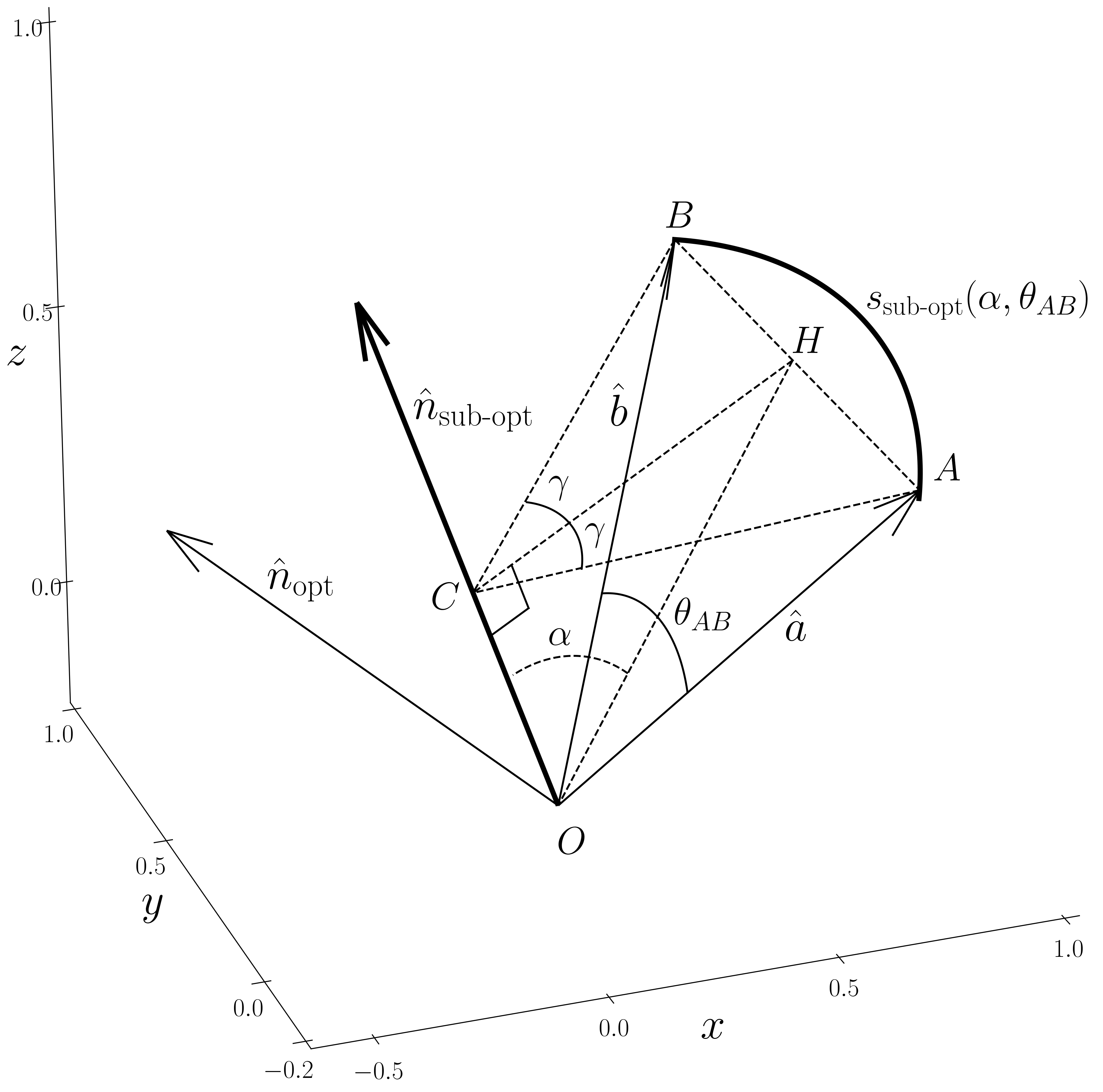}\caption{Schematic depiction of
the unit vectors $\hat{n}_{\mathrm{opt}}\protect\overset{\text{def}}{=}$
$(\hat{a}\times\hat{b})/\left\Vert \hat{a}\times\hat{b}\right\Vert $ and
$\hat{n}_{\mathrm{sub}\text{-}\mathrm{opt}}\protect\overset{\text{def}}{=}%
\cos(\alpha)\left[  (\hat{a}+\hat{b})/\left\Vert \hat{a}+\hat{b}\right\Vert
\right]  +\sin(\alpha)\left[  (\hat{a}\times\hat{b})/\left\Vert \hat{a}%
\times\hat{b}\right\Vert \right]  $ specifying the optimal (thin black) and
sub-optimal (thick black) rotation axes, respectively. The initial and final
Bloch vectors $\hat{a}$ and $\hat{b}$, respectively, are such that $\hat
{a}\cdot\hat{b}=\cos(\theta_{AB})$. The sub-optimal unitary evolution that
evolves $\hat{a}$ into $\hat{b}$ is specified by $\hat{n}_{\mathrm{sub}%
\text{-}\mathrm{opt}}$ and a rotation angle $\phi\protect\overset{\text{def}%
}{=}2\gamma$. Finally, the length of the nongeodesic path connecting these two
Bloch vectors via the sub-optimal unitary evolution is given by
$s_{\mathrm{sub}\text{-}\mathrm{opt}}\left(  \alpha\text{, }\theta
_{AB}\right)  \protect\overset{\text{def}}{=}\overline{AC}\left(
\alpha\text{, }\theta_{AB}\right)  \phi\left(  \alpha\text{, }\theta
_{AB}\right)  $ (thick black).}%
\end{figure}

\emph{Rotation axis} $\hat{n}_{\mathrm{sub}\text{\textrm{-}}\mathrm{opt}}$.
Let us begin with the construction of the rotation axis $\hat{n}%
_{\mathrm{sub}\text{\textrm{-}}\mathrm{opt}}$. Assume that $\rho
_{A}\overset{\text{def}}{=}\left\vert A\right\rangle \left\langle A\right\vert
=(\mathbf{1+}\hat{a}\cdot\mathbf{\boldsymbol{\sigma}})/2$ and $\rho
_{B}\overset{\text{def}}{=}\left\vert B\right\rangle \left\langle B\right\vert
=(\mathbf{1+}\hat{b}\cdot\mathbf{\boldsymbol{\sigma}})/2$. Since we wish to
arrive at $\hat{b}$ from $\hat{a}$ by means of a rotation, the path traced
will be a circumference embedded in the unit sphere $\hat{a}$ and $\hat{b}$
belong to. To define this rotation (a member of a larger family of rotations,
each one with a specific rotation axis and an angle of rotation), we need to
find the axis we have to rotate around and, in addition, the angle needed to
transition from $\hat{a}$ to $\hat{b}$. Any circumference embedded in a sphere
can be viewed as the intersection of a plane with the sphere. Since the
circumferences we focus on must contain the tips of $\hat{a}$ and $\hat{b}$,
such planes must contain the tips as well. This implies that the planes we
should take into consideration must contain the vector connecting $\hat{a}$
and $\hat{b}$. Then, the possible axes of rotations will correspond to unit
vectors with tail in the origin and, in addition, perpendicular to one of the
previously mentioned planes.

In particular, all possible axes of rotation belong to the same plane. The
plane can be defined as the two-dimensional vector space spanned by the set of
linearly independent vectors $\left\{  \hat{a}\times\hat{b}\text{, }\hat
{a}+\hat{b}\right\}  $. Note that this is a set of orthogonal non-unit
vectors. Moreover, any rotation axis can be defined by a unit vector $\hat
{n}_{\mathrm{sub}\text{\textrm{-}}\mathrm{opt}}$ written as a linear
combination of vectors in $\left\{  \hat{a}\times\hat{b}\text{, }\hat{a}%
+\hat{b}\right\}  $. Specifically, $\hat{n}_{\mathrm{sub}\text{\textrm{-}%
}\mathrm{opt}}$ can be expressed as%
\begin{equation}
\hat{n}_{\mathrm{sub}\text{\textrm{-}}\mathrm{opt}}\left(  \alpha\right)
\overset{\text{def}}{=}\cos\left(  \alpha\right)  \frac{\hat{a}+\hat{b}%
}{\left\Vert \hat{a}+\hat{b}\right\Vert }+\sin(\alpha)\frac{\hat{a}\times
\hat{b}}{\left\Vert \hat{a}\times\hat{b}\right\Vert }\text{,} \label{theshit}%
\end{equation}
with $\alpha\in\lbrack0$, $\pi]$. Assuming $\hat{a}\cdot\hat{b}=\cos\left(
\theta_{AB}\right)  $, $\hat{n}_{\mathrm{sub}\text{\textrm{-}}\mathrm{opt}%
}\left(  \alpha\right)  $ can be rewritten as%
\begin{equation}
\hat{n}_{\mathrm{sub}\text{\textrm{-}}\mathrm{opt}}\left(  \alpha\right)
=\cos\left(  \alpha\right)  \frac{\hat{a}+\hat{b}}{2\cos\left(  {\frac
{\theta_{AB}}{2}}\right)  }+\sin(\alpha)\frac{\hat{a}\times\hat{b}}%
{\sin\left(  {\theta}_{AB}\right)  }\text{.} \label{theshit1}%
\end{equation}
Eq. (\ref{theshit1}) expresses a one-parameter family of unit vectors that
determine rotations which bring $\hat{a}$ to $\hat{b}$. Interestingly, the
rotation axis $\hat{n}_{\mathrm{sub}\text{\textrm{-}}\mathrm{opt}}\left(
\alpha_{\ast}\right)  $ that is perpendicular to the plane containing $\hat
{a}$ and $\hat{b}$ is denoted as $\hat{n}_{\mathrm{opt}}\overset{\text{def}%
}{=}(\hat{a}\times\hat{b})/\sin\left(  {\theta}_{AB}\right)  $ for
$\alpha_{\ast}=\pi/2$. For completeness, we make explicit the fact that the
sub-optimal Hermitian Hamiltonian $\mathrm{H}_{\mathrm{sub}\text{-}%
\mathrm{opt}}$ is given by%
\begin{equation}
\mathrm{H}_{\mathrm{sub}\text{-}\mathrm{opt}}=\mathrm{H}_{\mathrm{sub}%
\text{-}\mathrm{opt}}\left(  \alpha\right)  \overset{\text{def}}{=}E\left[
\cos(\alpha)\frac{\hat{a}+\hat{b}}{\sqrt{(\hat{a}+\hat{b})\cdot(\hat{a}%
+\hat{b})}}+\sin(\alpha)\frac{\hat{a}\times\hat{b}}{\sqrt{(\hat{a}\times
\hat{b})\cdot(\hat{a}\times\hat{b})}}\right]  \cdot\mathbf{\boldsymbol{\sigma
}}\text{,} \label{SUB}%
\end{equation}
and, clearly, the unitary time propagator is $U_{\mathrm{sub}\text{-}%
\mathrm{opt}}(t)=\exp\left(  -\frac{i}{\hslash}\mathrm{H}_{\mathrm{sub}%
\text{-}\mathrm{opt}}t\right)  $.

\emph{Rotation angle} $\phi$. Let us recast the rotation angle as
$\phi=2\gamma$. From the drawing in Fig. $1$, consider the rectangular
triangle $AHC$ (where the angle $\angle$ $AHC$ is equal to $90^{0}$). Then,
from $\overline{CH}=\overline{AC}\cos\left(  \gamma\right)  $, we get
$\gamma=\arccos\left(  \overline{CH}/\overline{AC}\right)  $. The lengths of
the sides $\overline{CH}$ and $\overline{AC}$ can be expressed in terms of the
angles $\alpha$ and $\theta_{AB}$. From Fig. $1$, $\overline{CH}=\overline
{OH}\sin\left(  \alpha\right)  $, with $\overline{OH}=\left\Vert (\hat{a}%
+\hat{b})/2\right\Vert =\cos(\theta_{AB}/2)$. This is a consequence of the
fact that the triangle $OCH$ is rectangular, with the angle $\angle$ $OCH$
equal to $90^{0}$. Therefore, $\overline{CH}=$ $\sin\left(  \alpha\right)
\cos(\theta_{AB}/2)$. For clarity, observe that $\overline{OA}=\overline
{OB}=1$. Furthermore, since $\overline{AC}$ is the perpendicular component of
$\hat{a}$ with respect to $\hat{n}_{\mathrm{sub}\text{\textrm{-}}\mathrm{opt}%
}$, it can be expressed as the modulus of the cross product between $\hat{a}$
and $\hat{n}_{\mathrm{sub}\text{\textrm{-}}\mathrm{opt}}$. Specifically, we
have $\overline{AC}=\left\Vert \hat{n}_{\mathrm{sub}\text{\textrm{-}%
}\mathrm{opt}}\times\hat{a}\right\Vert =\sqrt{(\hat{n}_{\mathrm{sub}%
\text{\textrm{-}}\mathrm{opt}}\times\hat{a})\cdot(\hat{n}_{\mathrm{sub}%
\text{\textrm{-}}\mathrm{opt}}\times\hat{a})}$ with $\hat{a}=\left(  \hat
{a}\cdot\hat{n}_{\mathrm{sub}\text{\textrm{-}}\mathrm{opt}}\right)  \hat
{n}_{\mathrm{sub}\text{\textrm{-}}\mathrm{opt}}+\left[  \hat{a}-\left(
\hat{a}\cdot\hat{n}_{\mathrm{sub}\text{\textrm{-}}\mathrm{opt}}\right)
\hat{n}_{\mathrm{sub}\text{\textrm{-}}\mathrm{opt}}\right]  =\hat
{a}_{\parallel\text{, }\hat{n}_{\mathrm{sub}\text{\textrm{-}}\mathrm{opt}%
}\text{ }}+\hat{a}_{\bot\text{, }\hat{n}_{\mathrm{sub}\text{\textrm{-}%
}\mathrm{opt}}\text{ }}$. Therefore, $\overline{AC}=\left\Vert \hat
{n}_{\mathrm{sub}\text{\textrm{-}}\mathrm{opt}}\times\hat{a}_{\bot\text{,
}\hat{n}_{\mathrm{sub}\text{\textrm{-}}\mathrm{opt}}\text{ }}\right\Vert
=\left\Vert \hat{a}_{\bot\text{, }\hat{n}_{\mathrm{sub}\text{\textrm{-}%
}\mathrm{opt}}\text{ }}\right\Vert $. After some algebra and use of Eq.
(\ref{theshit1}), we get%
\begin{equation}
\overline{AC}=\overline{AC}\left(  \alpha\right)  \overset{\text{def}}{=}%
\sqrt{\frac{\cos^{2}({\alpha)}}{4\cos^{2}({\frac{\theta_{AB}}{2})}}\sin
^{2}(\theta_{AB}{)}+\sin^{2}({\alpha)}}\text{.} \label{boy}%
\end{equation}
Recalling that $\sin^{2}({\alpha)=1-\cos}^{2}\left(  \alpha\right)  $ and
$\sin^{2}(\theta_{AB}{)=4}\sin^{2}(\theta_{AB}/2{)\cos}^{2}(\theta_{AB}/2{)}$,
$\overline{AC}$ in Eq. (\ref{boy}) can be rewritten as%
\begin{equation}
\overline{AC}\left(  \alpha\right)  =\sqrt{1-\cos^{2}\left(  \alpha\right)
\cos^{2}\left(  \frac{\theta_{AB}}{2}\right)  }\text{.} \label{boyshit}%
\end{equation}
We emphasize that while Eq. (\ref{boy}) is derived using pure geometric
arguments and, in addition, the transition from Eq. (\ref{boy}) to Eq.
(\ref{boyshit}) relies on simple trigonometric manipulations, the more
succinct and geometric formula in Eq. (\ref{boyshit}) can be shown to
yield\textbf{ }$\overline{AC}\left(  \alpha\right)  =\sqrt{1-\cos^{2}\left(
\alpha\right)  \cos^{2}\left(  \theta_{AB}/2\right)  }=\Delta E\left(
\alpha\right)  /E$\textbf{ }thanks to quantum mechanics rules needed for the
calculation of the energy uncertainty of the system\textbf{ }$\Delta E\left(
\alpha\right)  $\textbf{. }For more details, we refer to Appendix A.\textbf{
}Finally, having expressed $\overline{CH}$ and $\overline{AC}$ in terms of
$\alpha$ and $\theta_{AB}$, the rotation angle $\phi=2\gamma=2\arccos\left(
\overline{CH}/\overline{AC}\right)  $ is given by
\begin{equation}
\phi=\phi\left(  \alpha\right)  \overset{\text{def}}{=}2\arccos{\left(
\frac{\sin({\alpha)}\cos({\frac{\theta_{AB}}{2})}}{\sqrt{1-\cos^{2}\left(
\alpha\right)  \cos^{2}\left(  \frac{\theta_{AB}}{2}\right)  }}\right)
=}2\arccos{\left(  \tan(\alpha)\frac{\left\Vert \hat{a}_{\parallel\text{,
}\hat{n}_{\mathrm{sub}\text{\textrm{-}}\mathrm{opt}}\text{ }}\right\Vert
}{\sqrt{1-\left\Vert \hat{a}_{\parallel\text{, }\hat{n}_{\mathrm{sub}%
\text{\textrm{-}}\mathrm{opt}}\text{ }}\right\Vert ^{2}}}\right)  }\text{.}
\label{yo}%
\end{equation}
In the last equality of Eq. (\ref{yo}), we expressed the rotation angle solely
in terms of\textbf{ }$\alpha$ and the magnitude of $\hat{a}_{\parallel\text{,
}\hat{n}_{\mathrm{sub}\text{\textrm{-}}\mathrm{opt}}\text{ }}%
\overset{\text{def}}{=}\left(  \hat{a}\cdot\hat{n}_{\mathrm{sub}%
\text{\textrm{-}}\mathrm{opt}}\right)  \hat{n}_{\mathrm{sub}\text{\textrm{-}%
}\mathrm{opt}}$. With the derivations of $\hat{n}_{\mathrm{sub}%
\text{\textrm{-}}\mathrm{opt}}\left(  \alpha\right)  $ in Eq. (\ref{theshit1})
and $\phi\left(  \alpha\right)  $ in Eq. (\ref{yo}), we can finally express in
an explicit manner $U_{\text{\textrm{sub-opt}}}(t_{AB})=e^{-i\frac{\phi}%
{2}\hat{n}_{\mathrm{sub}\text{\textrm{-}}\mathrm{opt}}\cdot\vec{\sigma}}$ with
$t_{AB}=t_{AB}\left(  \alpha\right)  \overset{\text{def}}{=}\left[
\hslash/(2E)\right]  \phi\left(  \alpha\right)  $.\begin{figure}[t]
\centering
\includegraphics[width=0.4\textwidth] {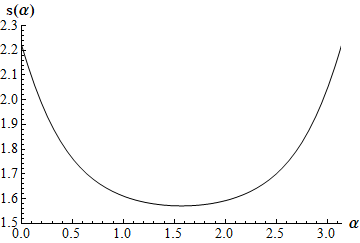}\caption{Plot of the length
$s=s\left(  \alpha\right)  \protect\overset{\text{def}}{=}s_{\mathrm{sub}%
\text{-}\mathrm{opt}}\left(  \alpha\text{, }\theta_{AB}\right)  $ of the path
generated by the Hamiltonian \textrm{H}$_{\mathrm{sub}\text{-}\mathrm{opt}%
}(\alpha)$ that connects two states $\left\vert A\right\rangle $ and
$\left\vert B\right\rangle $ with corresponding Bloch vectors $\hat{a}$ and
$\hat{b}$, respectively, with $\hat{a}\cdot\hat{b}=\cos(\theta_{AB})$. We set
$\theta_{AB}=\pi/2$ and visualize $s\left(  \alpha\right)  $ versus $\alpha$,
with $0\leq\alpha\leq\pi$. Note that the minimum $s$ is reached at $\alpha
=\pi/2\simeq1.57$ (i.e., when \textrm{H}$_{\mathrm{sub}\text{-}\mathrm{opt}%
}(\alpha)$ becomes \textrm{H}$_{\mathrm{opt}}$), with $s$ equal to $\pi/2$
(i.e., one-half of the shortest length $\pi$ between two orthogonal quantum
states on the Bloch sphere).}%
\end{figure}

\emph{Travel time and length of the path}. The travel time is simply given by
$t_{AB}=t_{AB}\left(  \alpha\right)  \overset{\text{def}}{=}\left[
\hslash/(2E)\right]  \phi\left(  \alpha\right)  $. Its explicit expression is%
\begin{equation}
t_{AB}\left(  \alpha\right)  =\frac{\hbar}{E}\arccos{\left(  \frac
{\sin({\alpha)}\cos({\frac{\theta_{AB}}{2})}}{\sqrt{1-\cos^{2}\left(
\alpha\right)  \cos^{2}\left(  \frac{\theta_{AB}}{2}\right)  }}\right)
}\text{.} \label{local}%
\end{equation}
Interestingly, we observe that the minimum of $t_{AB}\left(  \alpha\right)  $
in Eq. (\ref{local}) is achieved at $\alpha=\pi/2$, for any $\theta_{AB}$.
Since $\alpha=\pi/2$ implies $\hat{n}_{\mathrm{sub}\text{\textrm{-}%
}\mathrm{opt}}=\hat{n}_{\mathrm{opt}}$, the path is a geodesic since $\hat{a}$
rotates perpendicularly about $\hat{n}_{\mathrm{opt}}$. Indeed, as $\alpha$
approaches $\pi/2$, $t_{AB}\left(  \alpha\right)  $ approaches the optimal
(i.e., shortest) travel time given by $t_{AB}^{\mathrm{opt}}=\left[
\hslash/(2E)\right]  \theta_{AB}$. Moreover, the length $s$ of the arc arc
traced on the sphere during the motion is given by $s=s\left(  \alpha\right)
\overset{\text{def}}{=}\overline{AC}\left(  \alpha\right)  \phi\left(
\alpha\right)  $.\textbf{ }Explicitly, we have
\begin{equation}
s\left(  \alpha\right)  \overset{\text{def}}{=}2\arccos{\left(  \frac
{\sin({\alpha)}\cos({\frac{\theta_{AB}}{2})}}{\sqrt{1-\cos^{2}\left(
\alpha\right)  \cos^{2}\left(  \frac{\theta_{AB}}{2}\right)  }}\right)  }%
\cdot\sqrt{1-\cos^{2}\left(  \alpha\right)  \cos^{2}\left(  \frac{\theta_{AB}%
}{2}\right)  }\text{.} \label{yoyo}%
\end{equation}
As a simple cross check of the correctness of the expression for $s\left(
\alpha\right)  $ in Eq. (\ref{yoyo}), we consider the case of two antipodal
vectors $\hat{a}$ and $\hat{b}$. In this case, we have $\theta_{AB}=\pi$ and
we expect the length of the arc $s\left(  \alpha\right)  $ to be independent
from the particular axis of rotation and equal to $\pi$. This is justified by
the fact that any rotation transporting the initial state to the final state
will trace on the unit sphere an arc that corresponds to a semi-circumference
of unit radius. Indeed, it happens that $s\left(  \alpha\right)
\rightarrow\pi$ as $\theta_{AB}\rightarrow\pi$, for any choice of $\alpha$.
Moreover, for $\alpha=\pi/2$, we note that $s\left(  \alpha\right)  $ in Eq.
(\ref{yoyo}) reduces to its optimal (i.e., shortest) limiting value
$s_{\mathrm{opt}}=\theta_{AB}$. In our opinion, it is remarkable that
$s\left(  \alpha\right)  $ in Eq. (\ref{yoyo}) was obtained by means of simple
geometric arguments. Even more remarkable, is that we can explicitly show that
it coincides with the usual length of the path defined by Anandan and Aharonov
as
\begin{equation}
s\overset{\text{def}}{=}\int_{0}^{t_{AB}}2\frac{\Delta E(t)}{\hslash}%
dt=2\frac{\Delta E}{\hslash}t_{AB}\text{,} \label{sass}%
\end{equation}
with $\Delta E$ being the energy uncertainty defined as $\Delta
E\overset{\text{def}}{=}\left[  \left\langle A\left\vert \mathrm{H}%
_{\mathrm{sub}\text{\textrm{-}}\mathrm{opt}}^{2}\right\vert A\right\rangle
-\left\langle A\left\vert \mathrm{H}_{\mathrm{sub}\text{\textrm{-}%
}\mathrm{opt}}\right\vert A\right\rangle ^{2}\right]  ^{1/2}$. Interestingly,
we can verify that $\Delta E=\Delta E\left(  \alpha\right)
\overset{\text{def}}{=}E\sqrt{1-\cos^{2}\left(  \alpha\right)  \cos^{2}\left(
\frac{\theta_{AB}}{2}\right)  }=E\cdot\overline{AC}\left(  \alpha\right)  $.
For details on the calculation of $\Delta E\left(  \alpha\right)  $ with
ordinary (i.e., non-geometric) quantum mechanics rules, we refer to Appendix
A. Thus, recalling the expression of $t_{AB}\left(  \alpha\right)  $ in Eq.
(\ref{local}), it turns out that the \textquotedblleft
quantum\textquotedblright\ $s$ in Eq. (\ref{sass}) coincides with the
\textquotedblleft geometric\textquotedblright\ $s$ in Eq. (\ref{yoyo}). Using
Eqs. (\ref{boyshit}), (\ref{yoyo}) and (\ref{sass}), $s_{\mathrm{sub}%
\text{\textrm{-}}\mathrm{opt}}$ can be conveniently rewritten as%
\begin{equation}
s_{\mathrm{sub}\text{\textrm{-}}\mathrm{opt}}=s_{\mathrm{sub}\text{\textrm{-}%
}\mathrm{opt}}\left(  \alpha\right)  \overset{\text{def}}{=}2\sqrt{1-\cos
^{2}\left(  \alpha\right)  \cos^{2}\left(  \frac{\theta_{AB}}{2}\right)
}\arccos{\left[  \sin({\alpha)}\frac{\cos({\frac{\theta_{AB}}{2})}}%
{\sqrt{1-\cos^{2}\left(  \alpha\right)  \cos^{2}\left(  \frac{\theta_{AB}}%
{2}\right)  }}\right]  }\text{.}%
\end{equation}
In Fig. $3$, we show the behavior of $s\left(  \alpha\right)  $ as a function
of the parameter $\alpha$. Finally, we plot in Fig. $4$ the geodesic
efficiency $\eta_{\mathrm{GE}}\left(  \alpha\right)  \overset{\text{def}%
}{=}\theta_{AB}/s_{\mathrm{sub}\text{\textrm{-}}\mathrm{opt}}\left(
\alpha\right)  $ versus $\alpha$. We note that for $\theta_{AB}=\pi/2$ and
$\alpha$ very close to $\pi/2$, the approximate behavior of $\eta
_{\mathrm{GE}}\left(  \alpha\right)  $ is given by%
\begin{equation}
\eta_{\mathrm{GE}}\left(  \alpha\right)  =1-\frac{4-\pi}{4\pi}(\alpha
-\frac{\pi}{2})^{2}+O\left(  \alpha^{3}\right)  \text{.}%
\end{equation}
\begin{figure}[t]
\centering
\includegraphics[width=0.4\textwidth] {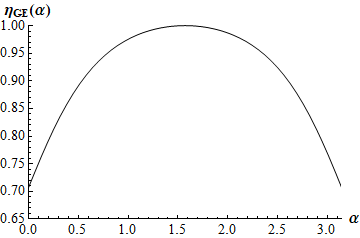}\caption{Plot of the geodesic
efficiency $\eta_{\mathrm{GE}}\left(  \alpha\right)  $ versus $\alpha$, with
$0\leq\alpha\leq\pi$. The sub-optimal quantum evolution specified by the
stationary Hamiltonian \textrm{H}$_{\mathrm{sub}\text{-}\mathrm{opt}}\left(
\alpha\right)  $ is assumed to connect $\left\vert A\right\rangle $ with
$\left\vert B\right\rangle $ with corresponding Bloch vectors $\hat{a}$ and
$\hat{b}$, respectively, where $\hat{a}\cdot\hat{b}=\cos\left(  \theta
_{AB}\right)  $. In the plot, we set $\theta_{AB}=\pi/2$. Observe that the
maximum geodesic efficiency value equal to one is obtained for $\alpha
=\pi/2\simeq1.57$, when \textrm{H}$_{\mathrm{sub}\text{-}\mathrm{opt}}\left(
\alpha\right)  $ reduces to the optimal Hamiltonian \textrm{H}$_{\mathrm{opt}%
}$.}%
\end{figure}

We are now ready to study both non-traceless and traceless time-varying qubit
Hamiltonians that lead to energy-wasteful quantum evolutions.

\subsection{From energy-resourceful to energy-wasteful evolutions}

Focusing on two-level quantum systems, we recall that in Ref. \cite{uzdin12} a
unit speed efficiency Hamiltonian evolution with $\eta_{\mathrm{SE}%
}\overset{\text{def}}{=}\Delta\mathrm{H}_{\rho}/\left\Vert \mathrm{H}%
\right\Vert _{\mathrm{SP}}=1$ is specified by $\mathrm{H}_{\mathrm{opt}%
}\left(  t\right)  \overset{\text{def}}{=}i\left\vert \partial_{t}%
m(t)\right\rangle \left\langle m(t)\right\vert -i\left\vert m(t)\right\rangle
\left\langle \partial_{t}m(t)\right\vert $ as in Eq. (\ref{optH}), where
$i\partial_{t}\left\vert m(t)\right\rangle =\mathrm{H}_{\mathrm{opt}}\left(
t\right)  \left\vert m(t)\right\rangle $ and $\mathrm{H}_{\mathrm{opt}}(t)$ is
a traceless Hermitian operator. Setting $\left\vert m\right\rangle =\left\vert
m(t)\right\rangle \overset{\text{def}}{=}c_{0}\left(  t\right)  \left\vert
0\right\rangle +c_{1}\left(  t\right)  \left\vert 1\right\rangle $ with
$\left\vert c_{0}\left(  t\right)  \right\vert ^{2}+\left\vert c_{1}\left(
t\right)  \right\vert ^{2}=1$, a straightforward calculation yields
$\Delta\mathrm{H}_{\rho}^{2}=\left\langle m\left\vert \mathrm{H}%
_{\mathrm{opt}}^{2}\right\vert m\right\rangle -\left\langle m\left\vert
\mathrm{H}_{\mathrm{opt}}\right\vert m\right\rangle ^{2}=\left\langle \dot
{m}\left\vert \dot{m}\right.  \right\rangle -0=\left\vert \dot{c}%
_{0}\right\vert ^{2}+\left\vert \dot{c}_{1}\right\vert ^{2}$ and $\left\Vert
\mathrm{H}\right\Vert _{\mathrm{SP}}^{2}=\left\vert \dot{c}_{0}\right\vert
^{2}+\left\vert \dot{c}_{1}\right\vert ^{2}$. Clearly, the dot denotes here
differentiation with respect to the time variable $t$. Therefore, the speed
efficiency becomes
\begin{equation}
\eta_{\mathrm{SE}}^{\mathrm{H}_{\mathrm{opt}}}(t)=\frac{\sqrt{\left\vert
\dot{c}_{0}\right\vert ^{2}+\left\vert \dot{c}_{1}\right\vert ^{2}}}%
{\sqrt{\left\vert \dot{c}_{0}\right\vert ^{2}+\left\vert \dot{c}%
_{1}\right\vert ^{2}}}=1 \label{J1A}%
\end{equation}
for any $t$ and, in addition, $\mathrm{H}_{\mathrm{opt}}\left(  t\right)  $
evolves the state $\left\vert m(t)\right\rangle $ with maximal speed and no
waste of energy resources. A class of suboptimal Hamiltonian evolutions that
generates the same path in projective Hilbert space can be specified by the
Hamiltonian given by \cite{uzdin12}%
\begin{equation}
\mathrm{H}_{\mathrm{sub}\text{-}\mathrm{opt}}^{\left(  \mathrm{trace}%
\neq0\right)  }(t)\overset{\text{def}}{=}\mathrm{H}_{\mathrm{opt}}\left(
t\right)  +\dot{\phi}\left\vert m\right\rangle \left\langle m\right\vert
\text{,} \label{optse1}%
\end{equation}
with the phase $\phi=\phi(t)\in%
\mathbb{R}
$ and $i\partial_{t}\left(  e^{-i\phi\left(  t\right)  }\left\vert
m(t)\right\rangle \right)  =\mathrm{H}_{\mathrm{sub}\text{-}\mathrm{opt}%
}^{\left(  \mathrm{trace}\neq0\right)  }\left(  e^{-i\phi\left(  t\right)
}\left\vert m(t)\right\rangle \right)  $. Setting $\left\vert m\right\rangle
=\left\vert m(t)\right\rangle \overset{\text{def}}{=}c_{0}\left(  t\right)
\left\vert 0\right\rangle +c_{1}\left(  t\right)  \left\vert 1\right\rangle $
with $\left\vert c_{0}\left(  t\right)  \right\vert ^{2}+\left\vert
c_{1}\left(  t\right)  \right\vert ^{2}=1$ and \textrm{H}$\overset{\text{def}%
}{=}\mathrm{H}_{\mathrm{sub}\text{-}\mathrm{opt}}^{\left(  \mathrm{trace}%
\neq0\right)  }$, a simple computation leads to $\Delta\mathrm{H}_{\rho}%
^{2}=\left\langle m\left\vert \mathrm{H}^{2}\right\vert m\right\rangle
-\left\langle m\left\vert \mathrm{H}\right\vert m\right\rangle ^{2}=\left(
\left\langle \dot{m}\left\vert \dot{m}\right.  \right\rangle +\dot{\phi}%
^{2}\right)  -\dot{\phi}^{2}=\left\vert \dot{c}_{0}\right\vert ^{2}+\left\vert
\dot{c}_{1}\right\vert ^{2}$ and $\left\Vert \mathrm{H}\right\Vert
_{\mathrm{SP}}^{2}=\dot{\phi}^{2}/2+\left\vert \dot{c}_{0}\right\vert
^{2}+\left\vert \dot{c}_{1}\right\vert ^{2}+(\left\vert \dot{\phi}\right\vert
/2)\sqrt{\dot{\phi}^{2}+4\left(  \left\vert \dot{c}_{0}\right\vert
^{2}+\left\vert \dot{c}_{1}\right\vert ^{2}\right)  }$. Therefore, the speed
efficiency reduces to%
\begin{equation}
\eta_{\mathrm{SE}}^{\mathrm{H}_{\mathrm{sub}\text{-}\mathrm{opt}}^{\left(
\mathrm{trace}\neq0\right)  }}(t)=\frac{\sqrt{\left\vert \dot{c}%
_{0}\right\vert ^{2}+\left\vert \dot{c}_{1}\right\vert ^{2}}}{\sqrt{\frac
{\dot{\phi}^{2}}{2}+\left(  \left\vert \dot{c}_{0}\right\vert ^{2}+\left\vert
\dot{c}_{1}\right\vert ^{2}\right)  +\frac{\left\vert \dot{\phi}\right\vert
}{2}\sqrt{\dot{\phi}^{2}+4\left(  \left\vert \dot{c}_{0}\right\vert
^{2}+\left\vert \dot{c}_{1}\right\vert ^{2}\right)  }}} \label{J1B}%
\end{equation}
with $0\leq\eta_{\mathrm{SE}}^{\mathrm{H}_{\mathrm{sub}\text{-}\mathrm{opt}%
}^{\left(  \mathrm{trace}\neq0\right)  }}(t)\leq1$. \ From Eqs. (\ref{J1A})
and (\ref{J1B}), the available energy resources increase thanks to phase
changes. Indeed, the size of $\mathrm{H}_{\mathrm{sub}\text{-}\mathrm{opt}%
}^{\left(  \mathrm{trace}\neq0\right)  }$ is greater than the size of
$\mathrm{H}_{\mathrm{opt}}$, $\left\Vert \mathrm{H}_{\mathrm{sub}%
\text{-}\mathrm{opt}}^{\left(  \mathrm{trace}\neq0\right)  }\right\Vert
_{\mathrm{SP}}\geq\left\Vert \mathrm{H}_{\mathrm{opt}}\right\Vert
_{\mathrm{SP}}$. Unfortunately, this increase in resources is wasted since the
speeds of quantum evolutions along the paths $t\mapsto\left\vert
m(t)\right\rangle $ and $t\mapsto e^{-i\phi\left(  t\right)  }\left\vert
m(t)\right\rangle $ corresponding to \textrm{H}$_{\mathrm{opt}}$ and
\textrm{H}$_{\mathrm{sub}\text{-}\mathrm{opt}}^{\left(  \mathrm{trace}%
\neq0\right)  }$, respectively, remain the same (i.e., $v_{\mathrm{H}%
_{\mathrm{opt}}}=v_{\mathrm{H}_{\mathrm{sub}\text{-}\mathrm{opt}}^{\left(
\mathrm{trace}\neq0\right)  }}$ since $\left(  \Delta\mathrm{H}_{\mathrm{opt}%
}\right)  _{\rho}=(\Delta\mathrm{H}_{\mathrm{sub}\text{-}\mathrm{opt}%
}^{\left(  \mathrm{trace}\neq0\right)  })_{\rho}$ with $\rho=\rho
(t)\overset{\text{def}}{=}\left\vert m(t)\right\rangle \left\langle
m(t)\right\vert $). In this sense, phase changes have a negative energetic
effect and should be avoided. Interestingly, in the event that phase changes
do occur, traceless sub-optimal Hamiltonians \textrm{H}$_{\mathrm{sub}%
\text{-}\mathrm{opt}}^{\left(  \mathrm{trace}=0\right)  }$ defined as
\begin{equation}
\mathrm{H}_{\mathrm{sub}\text{-}\mathrm{opt}}^{\left(  \mathrm{trace}%
=0\right)  }(t)\overset{\text{def}}{=}\mathrm{H}_{\mathrm{sub}\text{-}%
\mathrm{opt}}^{\left(  \mathrm{trace}\neq0\right)  }(t)-\frac{1}{2}%
\mathrm{tr}\left(  \dot{\phi}\left\vert m\right\rangle \left\langle
m\right\vert \right)  \mathbf{1=}\mathrm{H}_{\mathrm{sub}\text{-}\mathrm{opt}%
}^{\left(  \mathrm{trace}\neq0\right)  }(t)-\frac{1}{2}\dot{\phi}%
\mathbf{1}\text{,} \label{optse2}%
\end{equation}
with $\mathrm{H}_{\mathrm{sub}\text{-}\mathrm{opt}}^{\left(  \mathrm{trace}%
=0\right)  }(e^{-i\frac{\phi\left(  t\right)  }{2}}\left\vert m\right\rangle
)=i\partial_{t}(e^{-i\frac{\phi\left(  t\right)  }{2}}\left\vert
m\right\rangle )$, yield paths $t\mapsto e^{-i\frac{\phi\left(  t\right)  }%
{2}}\left\vert m\left(  t\right)  \right\rangle $ on the Bloch sphere that are
less energy wasteful than those of the nonzero trace sub-optimal Hamiltonians
\textrm{H}$_{\mathrm{sub}\text{-}\mathrm{opt}}^{\left(  \mathrm{trace}%
\neq0\right)  }(t)$ in Eq. (\ref{optse1}). For clarity, observe that
$\mathrm{tr}\left(  \dot{\phi}\left\vert m\right\rangle \left\langle
m\right\vert \right)  $ in Eq. (\ref{optse2}) is equal to \textrm{tr}%
$(\mathrm{H}_{\mathrm{sub}\text{-}\mathrm{opt}}^{\left(  \mathrm{trace}%
\neq0\right)  })$ since \textrm{tr}$(\mathrm{H}_{\mathrm{opt}})=0$. Traces can
be evaluated, for instance, with respect to the orthonormal basis $\left\{
\left\vert m\right\rangle \text{, }\left\vert \dot{m}\right\rangle
/\sqrt{\left\langle \dot{m}\left\vert \dot{m}\right.  \right\rangle }\right\}
$. Then, putting $\left\vert m\right\rangle =\left\vert m(t)\right\rangle
\overset{\text{def}}{=}c_{0}\left(  t\right)  \left\vert 0\right\rangle
+c_{1}\left(  t\right)  \left\vert 1\right\rangle $, with $\left\vert
c_{0}\left(  t\right)  \right\vert ^{2}+\left\vert c_{1}\left(  t\right)
\right\vert ^{2}=1$ and \textrm{H}$\overset{\text{def}}{=}\mathrm{H}%
_{\mathrm{sub}\text{-}\mathrm{opt}}^{\left(  \mathrm{trace}=0\right)  }$, an
easy calculation implies that $\Delta\mathrm{H}_{\rho}^{2}=\left\langle
m\left\vert \mathrm{H}^{2}\right\vert m\right\rangle -\left\langle m\left\vert
\mathrm{H}\right\vert m\right\rangle ^{2}=\left(  \left\langle \dot
{m}\left\vert \dot{m}\right.  \right\rangle +\dot{\phi}^{2}/4\right)
-\dot{\phi}^{2}/4=\left\vert \dot{c}_{0}\right\vert ^{2}+\left\vert \dot
{c}_{1}\right\vert ^{2}$ and $\left\Vert \mathrm{H}\right\Vert _{\mathrm{SP}%
}^{2}=\dot{\phi}^{2}/4+\left\vert \dot{c}_{0}\right\vert ^{2}+\left\vert
\dot{c}_{1}\right\vert ^{2}$. Therefore, the speed efficiency reduces to%
\begin{equation}
\eta_{\mathrm{SE}}^{\mathrm{H}_{\mathrm{sub}\text{-}\mathrm{opt}}^{\left(
\mathrm{trace}=0\right)  }}(t)=\frac{\sqrt{\left\vert \dot{c}_{0}\right\vert
^{2}+\left\vert \dot{c}_{1}\right\vert ^{2}}}{\sqrt{\frac{\dot{\phi}^{2}}%
{4}+\left(  \left\vert \dot{c}_{0}\right\vert ^{2}+\left\vert \dot{c}%
_{1}\right\vert ^{2}\right)  }}\text{,} \label{J1C}%
\end{equation}
with $0\leq\eta_{\mathrm{SE}}^{\mathrm{H}_{\mathrm{sub}\text{-}\mathrm{opt}%
}^{\left(  \mathrm{trace}\neq0\right)  }}(t)\leq\eta_{\mathrm{SE}}%
^{\mathrm{H}_{\mathrm{sub}\text{-}\mathrm{opt}}^{\left(  \mathrm{trace}%
=0\right)  }}(t)\leq\eta_{\mathrm{SE}}^{\mathrm{H}_{\mathrm{opt}}}(t)=1$ for
any $t$. It is worthwhile pointing out that when we set $\left\vert \dot
{c}_{0}\right\vert ^{2}+\left\vert \dot{c}_{1}\right\vert ^{2}=\mathcal{A}%
^{2}$ and assume $\dot{\phi}^{2}/(4\mathcal{A}^{2})\ll1$, the approximate
behavior of $\eta_{\mathrm{SE}}^{\mathrm{H}_{\mathrm{sub}\text{-}\mathrm{opt}%
}^{\left(  \mathrm{trace}=0\right)  }}$ in Eq. (\ref{J1C}) as a function of
$\dot{\phi}$ is given by%
\begin{equation}
\eta_{\mathrm{SE}}^{\mathrm{H}_{\mathrm{sub}\text{-}\mathrm{opt}}^{\left(
\mathrm{trace}=0\right)  }}\left(  \dot{\phi}\right)  =1-\frac{1}%
{8\mathcal{A}^{2}}\dot{\phi}^{2}+O(\dot{\phi}^{3})\text{.}%
\end{equation}
Moreover, from Eqs. (\ref{J1B}) and (\ref{J1C}), we note that although the
speed of quantum evolutions generated by $\mathrm{H}_{\mathrm{sub}%
\text{-}\mathrm{opt}}^{\left(  \mathrm{trace}\neq0\right)  }$ and
$\mathrm{H}_{\mathrm{sub}\text{-}\mathrm{opt}}^{\left(  \mathrm{trace}%
=0\right)  }$ are identical, the size of $\mathrm{H}_{\mathrm{sub}%
\text{-}\mathrm{opt}}^{\left(  \mathrm{trace}=0\right)  }$ is smaller than the
size of $\mathrm{H}_{\mathrm{sub}\text{-}\mathrm{opt}}^{\left(  \mathrm{trace}%
\neq0\right)  }$. Therefore, $\mathrm{H}_{\mathrm{sub}\text{-}\mathrm{opt}%
}^{\left(  \mathrm{trace}=0\right)  }$ yields paths on the Bloch sphere that
are less energy wasteful than those created by $\mathrm{H}_{\mathrm{sub}%
\text{-}\mathrm{opt}}^{\left(  \mathrm{trace}\neq0\right)  }$. To be clear,
although the paths are actually the same in projective Hilbert space in both
scenarios, in the more wasteful scenario a greater amount of energy is wasted
in the generation of global phases of the evolving quantum state since it is
not converted in (useful) speed of the quantum evolution. In Fig. $5$, we plot
$\eta_{\mathrm{SE}}(\dot{\phi})$ in Eq. (\ref{J1C}) versus $\dot{\phi}$ for a
suitable choice of $\left\vert m(t)\right\rangle $. Then, assuming to consider
the same $\left\vert m(t)\right\rangle $, we plot $\eta_{\mathrm{SE}}(t)$ in
Eq. (\ref{J1C}) versus $t$ for logarithmic, linear, and exponential temporal
growths of the phase $\phi\left(  t\right)  $ in Fig. $6$.\begin{figure}[t]
\centering
\includegraphics[width=0.4\textwidth] {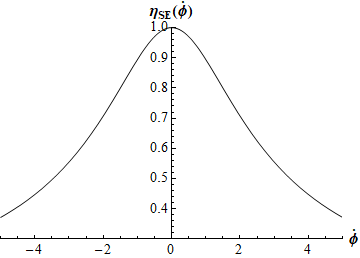}\caption{Plot of the speed
efficiency $\eta_{\mathrm{SE}}(\dot{\phi})$ in Eq. (\ref{J1C}) versus the rate
of change $\dot{\phi}$ of the phase $\phi$, with $-5\leq\dot{\phi}\leq5$ for
\textrm{H}$_{\mathrm{sub}\text{-}\mathrm{opt}}^{(\mathrm{trace}=0)}$. The
sub-optimal quantum evolution is specified by a nonstationary traceless
Hamiltonian \textrm{H}$_{\mathrm{sub}\text{-}\mathrm{opt}}^{(\mathrm{trace}%
=0)}$ described by a state $\left\vert m(t\right\rangle
\protect\overset{\text{def}}{=}\cos(\omega_{0}t)\left\vert 0\right\rangle
+\sin\left(  \omega_{0}t\right)  \left\vert 1\right\rangle $ and an arbitrary
phase $\phi=\phi\left(  t\right)  $. For simplicity, we set $\omega_{0}=1$ in
the plot. Physical units are specified by the choice of setting $\hslash=1$.}%
\end{figure}

For a final remark, let $\rho=\rho(t)\overset{\text{def}}{=}\left\vert
m(t)\right\rangle \left\langle m(t)\right\vert =\left[  \mathbf{1+}\hat
{a}\left(  t\right)  \cdot\mathbf{\boldsymbol{\sigma}}\right]  /2$. Then, it
is worthwhile pointing out that if we recast the Hamiltonians \textrm{H}%
$\left(  t\right)  =\mathrm{H}_{\mathrm{opt}}\left(  t\right)  $ along with
those in Eqs. (\ref{optse1}) and (\ref{optse2}) as \textrm{H}$\left(
t\right)  =\mathbf{h}_{\mathrm{opt}}\left(  t\right)  \cdot
\mathbf{\boldsymbol{\sigma}}$, $\mathrm{H}_{\mathrm{sub}\text{-}\mathrm{opt}%
}^{\left(  \mathrm{trace}\neq0\right)  }(t)=h_{0}\left(  t\right)
\mathbf{1+h}_{\mathrm{sub}\text{-}\mathrm{opt}}^{\left(  \mathrm{trace}%
\neq0\right)  }\left(  t\right)  \cdot\mathbf{\boldsymbol{\sigma}}$, and
$\mathrm{H}_{\mathrm{sub}\text{-}\mathrm{opt}}^{\left(  \mathrm{trace}%
=0\right)  }(t)=\mathbf{h}_{\mathrm{sub}\text{-}\mathrm{opt}}^{\left(
\mathrm{trace}=0\right)  }\left(  t\right)  \cdot\mathbf{\boldsymbol{\sigma}}%
$, respectively, we get $\hat{a}\left(  t\right)  \cdot\mathbf{h}%
_{\mathrm{opt}}\left(  t\right)  =0$, $h_{0}\left(  t\right)  +\hat{a}\left(
t\right)  \cdot\mathbf{h}_{\mathrm{sub}\text{-}\mathrm{opt}}^{\left(
\mathrm{trace}\neq0\right)  }\left(  t\right)  =\dot{\phi}\left(  t\right)
\neq0$, and $\hat{a}\left(  t\right)  \cdot\mathbf{h}_{\mathrm{sub}%
\text{-}\mathrm{opt}}^{\left(  \mathrm{trace}=0\right)  }\left(  t\right)
=\dot{\phi}\left(  t\right)  -(1/2)\mathrm{tr}\left(  \dot{\phi}\left\vert
m\right\rangle \left\langle m\right\vert \right)  \neq0$. Then, since
$\eta_{\mathrm{SE}}=\eta_{\mathrm{SE}}\left(  t\right)  \overset{\text{def}%
}{=}\sqrt{\mathbf{h}^{2}-(\hat{a}\cdot\mathbf{h})^{2}}/(\left\vert
h_{0}\right\vert +\sqrt{\mathbf{h}^{2}})$, traceless Hamiltonian evolutions
are not energy-wasteful only when $\hat{a}\cdot\mathbf{h=}0$. Instead, nonzero
trace Hamiltonian evolutions are always energy-wasteful. Summarizing, we
conclude that there is no waste of energy resources only for a quantum
evolution governed by a traceless Hamiltonian specified by a magnetic field
vector $\mathbf{h}_{\mathrm{opt}}\left(  t\right)  $ that is constantly
orthogonal to the the Bloch vector $\hat{a}\left(  t\right)  $. This could be
interpreted as \textquotedblleft instantaneous geodesic
motion\textquotedblright, since at every instant of time $\hat{a}\left(
t\right)  $ instantaneously rotates perpendicularly to $\mathbf{h}\left(
t\right)  $.

We are now ready to discuss some illustrative examples where we characterize
quantum evolutions in terms of geodesic, speed, and hybrid efficiency
measures.\begin{figure}[t]
\centering
\includegraphics[width=0.4\textwidth] {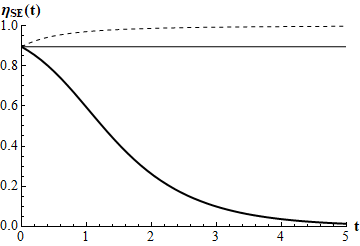}\caption{Plots of the speed
efficiency $\eta_{\mathrm{SE}}(t)$ in Eq. (\ref{J1C}) versus time $t$, with
$0\leq t\leq5$, for three scenarios specified by nonstationary traceless
Hamiltonians \textrm{H}$_{\mathrm{sub}\text{-}\mathrm{opt}}^{(\mathrm{trace}%
=0)}$ described by the state $\left\vert m(t\right\rangle
\protect\overset{\text{def}}{=}\cos(\omega_{0}t)\left\vert 0\right\rangle
+\sin\left(  \omega_{0}t\right)  \left\vert 1\right\rangle $ and phases
$\phi\left(  t\right)  \protect\overset{\text{def}}{=}\phi_{0}\ln\left[
1+\left(  \dot{\phi}_{0}/\phi_{0}\right)  t\right]  $ (dashed line),
$\phi\left(  t\right)  \protect\overset{\text{def}}{=}\phi_{0}+\dot{\phi}%
_{0}t$ (thin solid line), and $\phi\left(  t\right)
\protect\overset{\text{def}}{=}\phi_{0}+(e^{\dot{\phi}_{0}t}-1)$ (thick solid
line). Note that $\phi_{0}\protect\overset{\text{def}}{=}\phi\left(  0\right)
$ and $\dot{\phi}_{0}\protect\overset{\text{def}}{=}\dot{\phi}\left(
0\right)  $ in all three cases. For simplicity, we set in all three plots
$\omega_{0}=1$, $\phi_{0}=1$, and $\dot{\phi}_{0}=1$. Physical units are
determined so that $\hslash=1$.}%
\end{figure}

\section{Illustrative examples}

In this section, we exploit the formalism presented in the previous section to
study four illustrative examples. In particular, we display with these four
examples the full spectrum of possible quantum evolutions: Geodesic
unwasteful, geodesic wasteful, nongeodesic wasteful and, finally, nongeodesic
unwasteful evolutions. For each scenario, we characterize the average
properties of the quantum motion by means of the newly proposed hybrid
efficiency measure $\eta_{\mathrm{SE}}$ in Eq. (\ref{kelly}).

\subsection{First example: $(\eta_{\mathrm{GE}}$, $\eta_{\mathrm{SE}})=(1$,
$1)$}

To begin, we note that the stationary Hamiltonian \textrm{H}$_{\mathrm{opt}}%
=$\textrm{H}$_{\mathrm{sub}\text{\textrm{-}}\mathrm{opt}}\left(
\alpha\right)  $ with $\alpha=\pi/2$ and \textrm{H}$_{\mathrm{sub}%
\text{\textrm{-}}\mathrm{opt}}\left(  \alpha\right)  $ in Eq. (\ref{SUB})
yields quantum evolutions with $(\eta_{\mathrm{GE}}$, $\eta_{\mathrm{SE}}%
)=(1$, $1)$. However, we wish to present a time-varying setting here. In this
first example, we discuss a case in which the quantum evolution is specified
by a nonstationary Hamiltonian that yields an energy resourceful geodesic path
on the Bloch sphere. In other words, a path along which both the geodesic
efficiency $\eta_{\mathrm{GE}}$ and the speed efficiency $\eta_{\mathrm{SE}}$
are equal to one. We construct the Hamiltonian within the framework proposed
by Uzdin and collaborators in Ref. \cite{uzdin12}. Consider the following unit
state vector $\left\vert m\left(  t\right)  \right\rangle $ defined as,
\begin{equation}
\left\vert m\left(  t\right)  \right\rangle \overset{\text{def}}{=}\cos\left(
\frac{\theta\left(  t\right)  }{2}\right)  \left\vert 0\right\rangle
+e^{i\varphi_{0}}\sin\left(  \frac{\theta\left(  t\right)  }{2}\right)
\left\vert 1\right\rangle \text{,} \label{b}%
\end{equation}
with $\varphi_{0}\neq0$. Eq. (\ref{b}) represents a geodesic orbit since the
path on the Bloch sphere is a curve of constant longitude\textbf{ }%
$\varphi=\varphi_{0}$\textbf{ }circumnavigating the poles, with a speed that
varies with\textbf{ }$\dot{\theta}$. From Eq. (\ref{b}), we note that
$\left\langle m\left(  t\right)  \left\vert m\left(  t\right)  \right.
\right\rangle =1$ and $\left\langle \dot{m}\left(  t\right)  \left\vert
m\left(  t\right)  \right.  \right\rangle =0$. The unit Bloch vector $\hat
{a}\left(  t\right)  $ that corresponds to the pure state $\left\vert m\left(
t\right)  \right\rangle $ is given by $\hat{a}\left(  t\right)
\overset{\text{def}}{=}(\sin\left(  \theta\right)  \cos\left(  \varphi
_{0}\right)  $, $\sin(\theta)\sin\left(  \varphi_{0}\right)  $, $\cos
(\theta))$, with $\rho\left(  t\right)  \overset{\text{def}}{=}\left\vert
m\left(  t\right)  \right\rangle \left\langle m\left(  t\right)  \right\vert
=\left[  \mathbf{1}+\hat{a}\left(  t\right)  \cdot\mathbf{\boldsymbol{\sigma}%
}\right]  /2$. Given $\left\vert m\left(  t\right)  \right\rangle $ in Eq.
(\ref{b}), we get that the matrix representation of the traceless Hamiltonian
\textrm{H}$\left(  t\right)  \overset{\text{def}}{=}i(\left\vert \dot
{m}\right\rangle \left\langle m\right\vert -\left\vert m\right\rangle
\left\langle \dot{m}\right\vert )$ with respect to the computational basis
$\left\{  \left\vert 0\right\rangle \text{, }\left\vert 1\right\rangle
\right\}  $ is given by%
\begin{equation}
\mathrm{H}\left(  t\right)  =\left(
\begin{array}
[c]{cc}%
0 & -ie^{-i\varphi_{0}}\frac{\dot{\theta}}{2}\\
ie^{i\varphi_{0}}\frac{\dot{\theta}}{2} & 0
\end{array}
\right)  \text{,} \label{d}%
\end{equation}
where $\mathrm{H}\left(  t\right)  \left\vert m\left(  t\right)  \right\rangle
=i\partial_{t}\left\vert m\left(  t\right)  \right\rangle $. Then, setting
$\mathrm{H}\left(  t\right)  =\mathbf{h}\left(  t\right)  \cdot
\mathbf{\boldsymbol{\sigma}}$, we get%
\begin{equation}
\mathbf{h}\left(  t\right)  \overset{\text{def}}{=}\left(  -\frac{1}{2}%
\dot{\theta}\sin(\varphi_{0})\text{, }\frac{1}{2}\dot{\theta}\cos(\varphi
_{0})\text{, }0\right)  \text{.} \label{e}%
\end{equation}
Note that using $\hat{a}\left(  t\right)  \overset{\text{def}}{=}(\sin\left(
\theta\right)  \cos\left(  \varphi_{0}\right)  $, $\sin(\theta)\sin\left(
\varphi_{0}\right)  $, $\cos(\theta))$ and $\mathbf{h}\left(  t\right)  $ in
Eq. (\ref{e}), we correctly get that $\partial_{t}\hat{a}\left(  t\right)
=2\mathbf{h}\left(  t\right)  \times\hat{a}\left(  t\right)  $, $\forall t$.
From Eq. (\ref{e}), we have $\mathbf{h}\left(  t\right)  =h\left(  t\right)
\hat{h}(t)$ with a time-varying intensity $h\left(  t\right)  =\sqrt
{\dot{\theta}^{2}/4}$ and a time-constant direction $\hat{h}(t)=-\sin
(\varphi_{0})\hat{x}+\cos\left(  \varphi_{0}\right)  \hat{y}$ so that
$\partial_{t}\hat{h}(t)=\mathbf{0}$. Furthermore, $\mathbf{h}\left(  t\right)
$ is constantly perpendicular to the Bloch vector given that $\hat{a}\left(
t\right)  \cdot\mathbf{h}\left(  t\right)  =0$ for any $t$. From $\hat{a}%
\cdot\mathbf{h}=0$, we get $\partial_{t}\hat{a}\cdot\mathbf{h+}\hat{a}%
\cdot\mathbf{\dot{h}=}0$. However, since $\partial_{t}\hat{a}=2\mathbf{h}%
\times\hat{a}$ implies that $\partial_{t}\hat{a}\cdot\mathbf{h=}0$, we get
that $\hat{a}\cdot\mathbf{\dot{h}=}0$ when $\hat{a}\cdot\mathbf{h}=0$. Then,
since $\mathbf{h}$ does not change in direction, $\mathbf{h}$ and
$\mathbf{\dot{h}}$ are collinear. Given these geometric relations, it is worth
pointing out that the time-dependent curvature coefficient $\kappa
_{\mathrm{AC}}^{2}\left(  \hat{a}\text{, }\mathbf{h}\right)  $ given by
\cite{alsing24A,alsing24B}%
\begin{equation}
\kappa_{\mathrm{AC}}^{2}\left(  \mathbf{a}\text{, }\mathbf{h}\right)
\overset{\text{def}}{=}4\frac{\left(  \mathbf{a\cdot h}\right)  ^{2}%
}{\mathbf{h}^{2}-\left(  \mathbf{a\cdot h}\right)  ^{2}}+\frac{\left[
\mathbf{h}^{2}\mathbf{\dot{h}}^{2}-\left(  \mathbf{h\cdot\dot{h}}\right)
^{2}\right]  -\left[  \left(  \mathbf{a\cdot\dot{h}}\right)  \mathbf{h-}%
\left(  \mathbf{a\cdot h}\right)  \mathbf{\dot{h}}\right]  ^{2}}{\left[
\mathbf{h}^{2}-\left(  \mathbf{a\cdot h}\right)  ^{2}\right]  ^{3}}%
+4\frac{\left(  \mathbf{a\cdot h}\right)  \left[  \mathbf{a\cdot}\left(
\mathbf{h\times\dot{h}}\right)  \right]  }{\left[  \mathbf{h}^{2}-\left(
\mathbf{a\cdot h}\right)  ^{2}\right]  ^{2}}\text{,} \label{xxx}%
\end{equation}
with $\hat{a}=\mathbf{a}$ is identically zero and the evolution is geodesic.
The interested reader can find technical details leading to Eq. (\ref{xxx}) in
Appendix B. In Eq. (\ref{xxx}), \textquotedblleft\textrm{AC}\textquotedblright%
\ means Alsing and Cafaro. Alternatively, the geodesicity property can be
checked by verifying that the geodesic efficiency $\eta_{\mathrm{GE}}$ in Eq.
(\ref{efficiency}) is equal to one. This is indeed the case since,
essentially,%
\begin{equation}
\cos^{-1}\left[  \left\vert \left\langle m(t_{A}\left\vert m(t_{B}\right.
\right\rangle \right\vert \right]  =\int_{t_{A}}^{t_{B}}\sqrt{\left\langle
\dot{m}\left\vert \dot{m}\right.  \right\rangle }dt\text{,}%
\end{equation}
with $\cos^{-1}\left[  \left\vert \left\langle m(t_{A}\left\vert
m(t_{B}\right.  \right\rangle \right\vert \right]  =\left\vert \theta\left(
t_{B}\right)  -\theta\left(  t_{A}\right)  \right\vert /2=\int_{t_{A}}^{t_{B}%
}\sqrt{\left\langle \dot{m}\left\vert \dot{m}\right.  \right\rangle }dt$ given
that $\left\langle \dot{m}\left\vert \dot{m}\right.  \right\rangle
=\dot{\theta}^{2}/4$. Therefore, one can confirm that the evolution is a
geodesic one. Finally, by construction (i.e., given the expression of the
Hamiltonian \textrm{H}$\left(  t\right)  \overset{\text{def}}{=}i(\left\vert
\dot{m}\right\rangle \left\langle m\right\vert -\left\vert m\right\rangle
\left\langle \dot{m}\right\vert )$ as constructed in Ref. \cite{uzdin12}), the
evolution occurs with no waste of energy resources as well. In the
nonstationary quantum evolution considered here, we obtained a geodesic path
with a magnetic field $\mathbf{h}\left(  t\right)  $ in Eq. (\ref{e}) that
varies in magnitude and with a constant direction which is constantly
perpendicular to the Bloch vector $\hat{a}\left(  t\right)  $ of the evolving
quantum state $\left\vert m\left(  t\right)  \right\rangle $ in Eq. (\ref{b}).

At this point, at the cost of wasting energy resources with $\eta
_{\mathrm{SE}}<1$, we ask whether or not it is possible to have a vanishing
curvature of the quantum evolution in an alternative magnetic field
configuration where $\hat{a}\left(  t\right)  \cdot\mathbf{h}\left(  t\right)
\neq0$, $\partial_{t}\hat{h}(t)\neq\mathbf{0}$, and $\partial_{t}\hat
{a}\left(  t\right)  =2\mathbf{h}\left(  t\right)  \times\hat{a}\left(
t\right)  $ continues to be satisfied. In other words, is it possible to
generate a zero curvature and unit geodesic efficiency quantum evolution with
an applied magnetic field that changes both in magnitude and direction which,
in addition, is not constantly perpendicular to the Bloch vector of the
evolving quantum state? We address this question in the next example.

\subsection{Second example: $(\eta_{\mathrm{GE}}$, $\eta_{\mathrm{SE}})=(1$,
$<1)$}

In this second example, we assume to consider a quantum evolution governed by
a nonzero trace nonstationary Hamiltonian given by $\mathrm{H}\left(
t\right)  \overset{\text{def}}{=}i(\left\vert \dot{m}\right\rangle
\left\langle m\right\vert -\left\vert m\right\rangle \left\langle \dot
{m}\right\vert )+\dot{\phi}\left\vert m\right\rangle \left\langle m\right\vert
$ with a time-varying $\phi\left(  t\right)  \in%
\mathbb{R}
$ and $\left\vert m\left(  t\right)  \right\rangle $ as in Eq. (\ref{b}). From
Ref. \cite{uzdin12}, we know that the Hamiltonians in the first and second
example generate the same paths on the Bloch sphere. For this reason, the
Bloch vector in this second example remains $\hat{a}\left(  t\right)
\overset{\text{def}}{=}(\sin\left(  \theta\right)  \cos\left(  \varphi
_{0}\right)  $, $\sin(\theta)\sin\left(  \varphi_{0}\right)  $, $\cos
(\theta))$. The Hamiltonian $\mathrm{H}\left(  t\right)  $ is such that
$\mathrm{H}\left(  t\right)  (e^{-i\phi\left(  t\right)  }\left\vert
m(t)\right\rangle )=i\partial_{t}(e^{-i\phi\left(  t\right)  }\left\vert
m(t)\right\rangle )$. With respect to the computational basis $\left\{
\left\vert 0\right\rangle \text{, }\left\vert 1\right\rangle \right\}  $, the
matrix representation of $\mathrm{H}\left(  t\right)  $ becomes%
\begin{equation}
\mathrm{H}\left(  t\right)  =\left(
\begin{array}
[c]{cc}%
\dot{\phi}\cos^{2}\left(  \frac{\theta}{2}\right)  & \frac{e^{-i\varphi_{0}}%
}{2}\left[  -i\dot{\theta}+\dot{\phi}\sin(\theta)\right] \\
\frac{e^{i\varphi_{0}}}{2}\left[  i\dot{\theta}+\dot{\phi}\sin(\theta)\right]
& \dot{\phi}\sin^{2}\left(  \frac{\theta}{2}\right)
\end{array}
\right)  \text{.} \label{d2}%
\end{equation}
Then, setting $\mathrm{H}\left(  t\right)  =h_{0}\left(  t\right)
\mathbf{1+h}\left(  t\right)  \cdot\mathbf{\boldsymbol{\sigma}}$, we get that
$h_{0}(t)\overset{\text{def}}{=}\dot{\phi}/2$ and%
\begin{equation}
\mathbf{h}\left(  t\right)  \overset{\text{def}}{=}\left(
\begin{array}
[c]{c}%
\frac{1}{2}\cos\left(  \varphi_{0}\right)  \sin\left(  \theta\right)
\dot{\phi}-\frac{1}{2}\sin\left(  \varphi_{0}\right)  \dot{\theta}\\
\frac{1}{2}\sin\left(  \varphi_{0}\right)  \sin\left(  \theta\right)
\dot{\phi}+\frac{1}{2}\cos\left(  \varphi_{0}\right)  \dot{\theta}\\
\frac{1}{2}\cos(\theta)\dot{\phi}%
\end{array}
\right)  \text{.} \label{aa}%
\end{equation}
Note that when $\dot{\phi}=0$, $\mathbf{h}\left(  t\right)  $ in Eq.
(\ref{aa}) reduces to $\mathbf{h}\left(  t\right)  $ in Eq. (\ref{e}).
Furthermore, it can be explicitly checked that we correctly obtain that
$\partial_{t}\hat{a}\left(  t\right)  =2\mathbf{h}\left(  t\right)  \left(
t\right)  \times\hat{a}\left(  t\right)  $ where $\hat{a}\left(  t\right)
\overset{\text{def}}{=}(\sin\left(  \theta\right)  \cos\left(  \varphi
_{0}\right)  $, $\sin(\theta)\sin\left(  \varphi_{0}\right)  $, $\cos
(\theta))$ and $\mathbf{h}\left(  t\right)  $ as in Eq. (\ref{aa}). Observe
that in this second example, we have $\hat{a}\left(  t\right)  \cdot
\mathbf{h}\left(  t\right)  =\dot{\phi}/2\neq0$, $\mathbf{h}^{2}\left(
t\right)  =(\dot{\phi}^{2}+\dot{\theta}^{2})/4$ is time-varying, and\textbf{
}$\partial_{t}\hat{h}\left(  t\right)  \neq\mathbf{0}$. For simplicity of
calculations, we assume in what follows that $\varphi_{0}=0$, $\phi\left(
t\right)  =\phi_{0}+\Omega_{0}t$, and $\theta\left(  t\right)  =\theta
_{0}+\omega_{0}t$. This way, we have $\dot{\phi}=\Omega_{0}$ and $\dot{\theta
}=\omega_{0}$. In these working assumptions, we have $\hat{a}\left(  t\right)
=\mathbf{a}(t)=\left(  \sin\left(  \theta\right)  \text{, }0\text{, }%
\cos\left(  \theta\right)  \right)  $, $\mathbf{h}\left(  t\right)  =\left(
\frac{\Omega_{0}}{2}\sin\left(  \theta\right)  \text{, }\frac{\omega_{0}}%
{2}\text{, }\frac{\Omega_{0}}{2}\cos\left(  \theta\right)  \right)  $, and
$\mathbf{\dot{h}}\left(  t\right)  =\left(  \frac{\Omega_{0}\omega_{0}}{2}%
\cos\left(  \theta\right)  \text{, }0\text{, }-\frac{\Omega_{0}\omega_{0}}%
{2}\sin\left(  \theta\right)  \right)  $. Then, a straightforward but tedious
calculation yields $\kappa_{\mathrm{AC}}^{2}\left(  \mathbf{a}\text{,
}\mathbf{h}\right)  =0$, with the curvature coefficient defined in Eq.
(\ref{xxx}). This was expected since the evolution paths are the same as in
the first example. Nevertheless, we checked the vanishing of the curvature
coefficient to double-check the correctness of the expression in Eq.
(\ref{xxx}). For consistency check, one can also verify that $\eta
_{\mathrm{GE}}=1$ in this second case as well. Interestingly, recalling that
only $h_{\perp}\left(  t\right)  $ plays a key role in the calculation of the
geodesic efficiency $\eta_{\mathrm{GE}}$ along with the fact that $\mathbf{h}$
can be decomposed as $\mathbf{h=h}_{\parallel}\mathbf{+h}_{\perp}%
\mathbf{=}\left(  \mathbf{h\cdot}\hat{a}\right)  \hat{a}+\left[
\mathbf{h-}\left(  \mathbf{h\cdot}\hat{a}\right)  \hat{a}\right]  $, we
observe that $\left(  \mathbf{h}_{\perp}\right)  _{\mathrm{case}%
\text{\textrm{-}}\mathrm{2}}=\left(  \mathbf{h}_{\perp}\right)
_{\mathrm{case}\text{\textrm{-}}\mathrm{1}}=\left(  -(1/2)\sin(\varphi
_{0})\dot{\theta}\text{, }(1/2)\cos(\varphi_{0})\dot{\theta}\text{, }0\right)
$. Finally, unlike the first example, we have $\eta_{\mathrm{SE}}\left(
t\right)  =h_{\perp}\left(  t\right)  /\left[  \left\vert h_{0}\left(
t\right)  \right\vert +\sqrt{h_{\parallel}^{2}(t)+h_{\perp}^{2}\left(
t\right)  }\right]  <1$. Specifically, it happens that $h_{\perp}^{2}\left(
t\right)  =\dot{\theta}^{2}/4$, $h_{\parallel}^{2}(t)=\dot{\phi}^{2}/4$, and
$h_{0}\left(  t\right)  =\dot{\phi}/2$. Therefore, unlike the first example,
this second example is specified by a nonzero trace Hamiltonian with
$h_{\parallel}(t)\neq0$. The departure from the traceless condition together
with the emergence of a component of the magnetic field along the Bloch vector
$\hat{a}\left(  t\right)  $ lead to a quantum evolution in which energy
resources are wasted, despite being a geodesic evolution.

In summary, in these first two examples, we have used Uzdin's formalism to
show two main points. First, unlike in a stationary setting, in a
nonstationary setting it is not necessary to have $\hat{a}\left(  t\right)
\cdot\mathbf{h}\left(  t\right)  =0$ for having a geodesic evolution. Second,
a geodesic path can be energetically wasteful. In this next pair of examples,
we show that it is not sufficient to have $\hat{a}\left(  t\right)
\cdot\mathbf{h}\left(  t\right)  =0$ in a nonstationary setting in order to
produce a geodesic quantum evolution. Moreover, we shall see that given a
wasteful Hamiltonian specified by a magnetic field with a high intensity, it
is possible to create a new Hamiltonian that generates the same paths on the
Bloch sphere as the original Hamiltonian, while mitigating the energy waste by
removing unuseful\textbf{ }parts of the original magnetic field and lowering
its intensity (in an effort to drive $h_{\Vert}\rightarrow0$ ). We are now
ready to discuss the next two examples.

\subsection{Third example: $(\eta_{\mathrm{GE}}$, $\eta_{\mathrm{SE}})=(<1$,
$<1)$}

To begin, we observe that the constant Hamiltonians \textrm{H}$_{\mathrm{sub}%
\text{\textrm{-}}\mathrm{opt}}\left(  \alpha\right)  $ with $\alpha\neq\pi/2$
in Eq. (\ref{SUB}) lead to quantum evolutions with $(\eta_{\mathrm{GE}}$,
$\eta_{\mathrm{SE}})=(<1$, $<1)$. In particular, given $\Delta E\left(
\alpha\right)  $ calculated in Appendix A, the speed efficiency in Eq.
(\ref{se1}) reduces to $\eta_{\mathrm{SE}}\left(  \alpha\right)  =\Delta
E\left(  \alpha\right)  /E=\sqrt{1-\cos^{2}(\alpha)\cos^{2}(\theta_{AB}/2)}$.
For this reason, we consider in this third example a quantum evolution
specified by a time-independent Hamiltonian given by \textrm{H}%
$\overset{\text{def}}{=}\mathbf{h\cdot\boldsymbol{\sigma}=}\gamma\sigma_{z}$,
with $\mathbf{h=}\gamma\hat{z}$ and $\gamma>0$. Furthermore, we assume that
the initial state is given by $\left\vert A\right\rangle \overset{\text{def}%
}{=}(\sqrt{3}/2)|0\rangle+(1/2)|1\rangle$ with $\rho_{A}\overset{\text{def}%
}{=}\left\vert A\right\rangle \left\langle A\right\vert =(\mathbf{1+}\hat
{a}\cdot\mathbf{\boldsymbol{\sigma}})/2$. The unit Bloch vector equals
$\hat{a}\overset{\text{def}}{=}(\sqrt{3}/2)\hat{x}+(1/2)\hat{z}$ and is such
that $\hat{a}\cdot\mathbf{h\neq}0$. The action\textbf{ }of the unitary time
propagator $U(t)=e^{-i\mathrm{H}t}$ on $\left\vert A\right\rangle $ yields
\begin{equation}
|\psi(t)\rangle=e^{-i\gamma t}\left[  \frac{\sqrt{3}}{2}|0\rangle+\frac{1}%
{2}e^{i2\gamma t}|1\rangle\right]  \text{.} \label{leo1}%
\end{equation}
We can notice from Eq. (\ref{leo1}) that, apart for a global phase factor that
is not relevant when considering the evolution on the Bloch sphere, the only
parameter that changes is the azimuthal angle $\varphi$ according to
$\varphi=\varphi(t)\overset{\text{def}}{=}2\gamma t$. This is in agreement
with the fact that the operator $e^{-i\gamma\sigma_{z}t}$ corresponds to a
rotation of angle $2\gamma t$ around the axis $\hat{z}$ in the Bloch sphere. A
straightforward calculation yields $\kappa_{\mathrm{AC}}^{2}\left(
\mathbf{a}\text{, }\mathbf{h}\right)  \overset{\text{def}}{=}4\left(
\mathbf{a\cdot h}\right)  ^{2}/\left[  \mathbf{h}^{2}-\left(  \mathbf{a\cdot
h}\right)  ^{2}\right]  =4/3\neq0$ and $\eta_{\mathrm{GE}}\overset{\text{def}%
}{=}s_{0}/s\leq1$ evaluated between $\left\vert A\right\rangle $ and
$\left\vert \psi(t)\right\rangle $ given by,%
\begin{equation}
\eta_{\mathrm{GE}}=\frac{2}{\sqrt{3}}\frac{\arccos\left(  \frac{\sqrt{\left[
3+\cos(2\gamma t)\right]  ^{2}+\sin^{2}(2\gamma t)}}{4}\right)  }{\gamma
t}\text{.}%
\end{equation}
Therefore, this quantum evolution is nongeodesic. Furthermore, we can
explicitly verify that the speed efficiency $\eta_{\mathrm{SE}}%
\overset{\text{def}}{=}\Delta\mathrm{H}_{\rho}/\left\Vert \mathrm{H}%
\right\Vert _{\mathrm{SP}}$ of this evolution is smaller than $1$. Indeed,
since $\Delta\mathrm{H}_{\rho}^{2}=(3/4)\gamma^{2}$ and $\left\Vert
\mathrm{H}\right\Vert _{\mathrm{SP}}^{2}=\gamma^{2}$, we find $\eta
_{\mathrm{SE}}=\sqrt{3}/2<1$. In conclusion, the quantum evolution considered
is neither geodesic nor energy-resourceful.

In our final example, starting from the time-independent Hamiltonian of this
third example, we construct a new time-dependent Hamiltonian that generates
the same evolution paths on the Bloch sphere (i.e., same nonvanishing
curvature coefficient $\kappa_{\mathrm{AC}}^{2}$ and same non-unit geodesic
efficiency $\eta_{\mathrm{GE}}$), but with no waste of energy resources (i.e.,
$\eta_{\mathrm{SE}}=1$).\begin{table}[t]
\centering
\begin{tabular}
[c]{c|c|c|c|c}\hline\hline
\textbf{Example} & $\bar{\eta}_{\mathrm{GE}}$ & $\bar{\eta}_{\mathrm{SE}}$ &
$\eta_{\mathrm{HE}}$ & \textbf{Type of quantum evolution}\\\hline
First & $1$ & $1$ & $1$ & Geodesic unwasteful\\\hline
Second & $1$ & $\sim0.87$ & $\sim0.87$ & Geodesic wasteful\\\hline
Third & $\sim0.98$ & $\sqrt{3}/2$ & $\sim0.85$ & Nongeodesic wasteful\\\hline
Fourth & $\sim0.98$ & $1$ & $\sim0.98$ & Nongeodesic unwasteful\\\hline
\end{tabular}
\caption{Numerical estimates of the average geodesic efficiency, average speed
efficiency, and hybrid efficiency. Time-averages are estimated over the unit
time interval. In the second example, we assume $\theta\left(  t\right)
\protect\overset{\text{def}}{=}\omega_{0}t$ and $\phi\left(  t\right)
\protect\overset{\text{def}}{=}\nu_{0}t$. For simplicity, we set $\omega
_{0}=1$ and $\nu_{0}=10^{-1}$ in the numerical calculations. Moreover, in the
third and fourth examples, we put $\gamma=1$. Finally, physical units are
determined so that the reduced Planck constant can be set equal to one.}%
\end{table}

\subsection{Fourth example: $(\eta_{\mathrm{GE}}$, $\eta_{\mathrm{SE}})=(<1$,
$1)$}

We construct here a traceless time-dependent Hamiltonian with $100\%$ speed
efficiency following Uzdin's prescription. The Hamiltonian is given by
$\mathrm{H}\left(  t\right)  =i\left(  |\partial_{t}m\rangle\langle
m|-|m\rangle\langle\partial_{t}m|\right)  $, where $|m\rangle$ is the parallel
transported vector defined in terms of $|\psi(t)\rangle$ in Eq. (\ref{leo1})
as
\begin{equation}
|m(t)\rangle\overset{\text{def}}{=}e^{-\int_{0}^{t}\langle\psi(t^{\prime
})|\partial_{t^{\prime}}\psi(t^{\prime})\rangle dt^{\prime}}|\psi
(t)\rangle\text{,} \label{145}%
\end{equation}
with $\partial_{t^{\prime}}\overset{\text{def}}{=}\partial/\partial t^{\prime
}$. Exploiting the time-dependent Schr\"{o}dinger evolution equation
$i\partial_{t}|\psi(t)\rangle=\mathrm{H}\left(  t\right)  |\psi(t)\rangle$ to
evaluate the integral in Eq. (\ref{145}), $|m(t)\rangle$ becomes
\begin{equation}
|m(t)\rangle=\frac{\sqrt{3}}{2}e^{-i\frac{\gamma}{2}t}|0\rangle+\frac{1}%
{2}e^{i\frac{3\gamma}{2}t}|1\rangle\text{.} \label{147}%
\end{equation}
Note that from Eq. (\ref{147}) we could write $|m(t)\rangle_{\mathrm{case}%
\text{-}4}=e^{i\frac{\gamma}{2}t}|m(t)\rangle_{\mathrm{case}\text{-}3}$, using
Eq. (\ref{leo1}) on the right-hand side. From Eq. (\ref{147}), we can easily
check that $\langle m(t)|m(t)\rangle=1$ and $\langle m(t)|\partial
_{t}m(t)\rangle=0$. Using Eq. (\ref{147}), $\mathrm{H}\left(  t\right)  $ can
be recast as%
\begin{equation}
\mathrm{H}\left(  t\right)  =\frac{3}{4}\gamma|0\rangle\langle0|-\frac
{\sqrt{3}}{4}\gamma e^{i2\gamma t}|1\rangle\langle0|-\frac{\sqrt{3}}{4}\gamma
e^{-i2\gamma t}|0\rangle\langle1|-\frac{3}{4}\gamma|1\rangle\langle1|\text{.}
\label{148}%
\end{equation}
In terms of the expression \textrm{H}$(t)=\mathbf{h(}t\mathbf{)\cdot
\boldsymbol{\sigma}}$, $\mathrm{H}\left(  t\right)  $ in Eq. (\ref{148}) is
specified by $\mathbf{h(}t\mathbf{)}\overset{\text{def}}{\mathbf{=}}%
\frac{\sqrt{3}}{2}\gamma\hat{h}(t)$, with $\hat{h}(t)$ given by%
\begin{equation}
\hat{h}(t)\overset{\text{def}}{=}\left(  -\frac{1}{2}\cos{\left(  2\gamma
t\right)  }\text{, }-\frac{1}{2}\sin{\left(  2\gamma t\right)  }\text{, }%
\frac{\sqrt{3}}{2}\right)  \text{.} \label{153}%
\end{equation}
Comparing $\left(  \mathbf{h}\right)  _{\mathrm{case}\text{-}3}%
\overset{\text{def}}{\mathbf{=}}\gamma\hat{z}$ with $\left(  \mathbf{h}%
\right)  _{\mathrm{case}\text{-}4}\overset{\text{def}}{\mathbf{=}}(\sqrt
{3}/2)\gamma\hat{h}(t)$, we note that the magnitude of the magnetic field in
case-$4$ is smaller than the magnitude of the magnetic field in case-$3$.
Interestingly, we note that the time-dependent unit Bloch vector $\hat
{a}\left(  t\right)  $ that corresponds to the states $\left\vert \psi\left(
t\right)  \right\rangle $ and $\left\vert m(t)\right\rangle $ in Eqs.
(\ref{leo1}) and (\ref{147}), respectively, is given by
\begin{equation}
\hat{a}\left(  t\right)  \overset{\text{def}}{=}\left(  \frac{\sqrt{3}}{2}%
\cos{\left(  2\gamma t\right)  }\text{, }\frac{\sqrt{3}}{2}\sin{\left(
2\gamma t\right)  }\text{, }\frac{1}{2}\right)  \text{.} \label{154}%
\end{equation}
From Eqs. (\ref{153}) and (\ref{154}), it is straightforward to verify that
$\hat{a}\left(  t\right)  \perp\hat{h}(t)$. Therefore, since $\mathbf{h}$ can
be decomposed as $\mathbf{h=h}_{\parallel}\mathbf{+h}_{\perp}\mathbf{=}\left(
\mathbf{h\cdot}\hat{a}\right)  \hat{a}+\left[  \mathbf{h-}\left(
\mathbf{h\cdot}\hat{a}\right)  \hat{a}\right]  $ and given that $\eta
_{\mathrm{SE}}\left(  t\right)  =h_{\perp}\left(  t\right)  /\sqrt
{h_{\parallel}^{2}(t)+h_{\perp}^{2}\left(  t\right)  }$, we have in our case
here that $h_{\parallel}(t)=0$ and $\eta_{\mathrm{SE}}\left(  t\right)  $ is
identically equal to one. In conclusion, we can interpret the waste in energy
resources in terms of the presence of useless parallel magnetic field when
comparing case-$3$ with case-$4$. Finally, given $\mathbf{h(}t\mathbf{)}%
\overset{\text{def}}{\mathbf{=}}\frac{\sqrt{3}}{2}\gamma\hat{h}(t)$ and
$\hat{a}\left(  t\right)  $ in Eq. (\ref{154}), we note that $\hat{a}\left(
t\right)  \cdot\mathbf{h(}t\mathbf{)=}0$, $\hat{a}\left(  t\right)
\cdot\mathbf{\dot{h}(}t\mathbf{)=}0$, and $\mathbf{h(}t\mathbf{)\cdot\dot{h}%
(}t\mathbf{)=}0$. Therefore, the cumbersome expression of the curvature
coefficient in Eq. (\ref{xxx})\ reduces to $\kappa_{\mathrm{AC}}^{2}\left(
\mathbf{a}\text{, }\mathbf{h}\right)  =\mathbf{\dot{h}}^{2}\mathbf{(}%
t\mathbf{)/h}^{4}\mathbf{(}t\mathbf{)=}4/3$. This result coincides, as
expected, with the result obtained in the third example presented earlier. As
a final remark, based on what we have shown in this example, we stress that in
a nonstationary setting (unlike the stationary setting), the condition
$\hat{a}\left(  t\right)  \cdot\mathbf{h}\left(  t\right)  =0$ is not
generally sufficient to yield a geodesic quantum evolution. It becomes
sufficient only when $\mathbf{h}$ and $\mathbf{\dot{h}}$ are collinear (i.e.,
when $\mathbf{h}$ does not change in direction). This can also be seen in a
very neat manner as follows. When $\mathbf{a}\cdot\mathbf{h}=0$ (with
$\mathbf{a=}\hat{a}$), $\mathbf{\dot{a}}\cdot\mathbf{h+a}\cdot\mathbf{\dot{h}%
}=0$. Then, since $\mathbf{\dot{a}=2h\times a}$ implies $\mathbf{\dot{a}\cdot
h=}0$, we have $\mathbf{a}\cdot\mathbf{\dot{h}}=0$. Then, $\kappa
_{\mathrm{AC}}^{2}\left(  \mathbf{a}\text{, }\mathbf{h}\right)  $ in Eq.
(\ref{xxx}) reduces to%
\begin{equation}
\kappa_{\mathrm{AC}}^{2}\left(  \mathbf{a}\text{, }\mathbf{h}\right)
=\frac{\mathbf{h}^{2}\mathbf{\dot{h}}^{2}-\left(  \mathbf{h\cdot\dot{h}%
}\right)  ^{2}}{\left[  \mathbf{h}^{2}-\left(  \mathbf{a\cdot h}\right)
^{2}\right]  ^{3}}\text{.} \label{c4}%
\end{equation}
Setting $\mathbf{h=h}_{\bot}=h_{\perp}\left(  t\right)  \hat{h}_{\perp}(t)$
(since $\mathbf{h}_{\parallel}=\mathbf{0}$, given that $\mathbf{a}%
\cdot\mathbf{h}=0$), after some algebra, Eq. (\ref{c4}) can be recast as%
\begin{equation}
\kappa_{\mathrm{AC}}^{2}\left(  t\right)  =\frac{\frac{d\hat{h}_{\perp}}%
{dt}\cdot\frac{d\hat{h}_{\perp}}{dt}}{\mathbf{h}_{\bot}^{2}}\text{.}
\label{c5}%
\end{equation}
From Eq. (\ref{c5}), it is evident that $\kappa_{\mathrm{AC}}^{2}\left(
t\right)  $ vanishes in a time-dependent setting with $\mathbf{a}%
\cdot\mathbf{h}=0$ if and only if the magnetic field does not change in
direction. With this remark, we end our presentation of these four
illustrative examples. A summary of the numerical estimates of the average
geodesic efficiency, average speed efficiency, and hybrid efficiency for each
one of the four examples considered here are presented in Table II.

We are now ready for our summary of results and final considerations.

\section{Concluding remarks}

In this paper, we investigated different families of sub-optimal qubit
Hamiltonians, both stationary (Eq.(\ref{SUB})) and time-varying (Eqs.
(\ref{optse1}) and (\ref{optse2})), for which the so-called geodesic
efficiency (Eq. (\ref{efficiency})) and the speed efficiency (Eq. (\ref{se1}))
of the corresponding quantum evolutions are less than one. The detrimental
effects caused by the stationary and nonstationary Hamiltonians and quantified
by means of the geodesic and speed efficiency, respectively, are illustrated
in Figs. $3$ and $4$ (stationary case) and Figs. $5$ and $6$ (nonstationary
case). In addition, we proposed a different hybrid efficiency measure (Eq.
(\ref{kelly})) that combines the two efficiency measures mentioned earlier. As
reported in Table I, this hybrid efficiency quantifier allows us to categorize
quantum evolutions into four groups: Geodesic unwasteful, nongeodesic
unwasteful, geodesic wasteful, and nongeodesic wasteful. Then, for each and
everyone of these groups, we exhibited practical illustrative examples that
demonstrate how this hybrid measure captures the overall deviations from
optimal timing and perfect speed efficiency within a specific time-frame. As
summary of results appears in Table II. Ultimately, after considering the
notion of curvature in quantum evolution (Eq. (\ref{xxx})), we explored
Hamiltonians that are determined by magnetic field arrangements, whether they
are stationary or nonstationary. These Hamiltonians result in an optimal
hybrid efficiency, incorporating both time-optimality and $100\%$ speed
efficiency, within a finite time period.

\medskip

The main conclusions that can be derived are as follows. First\textbf{,} the
geodesic efficiency $\eta_{\mathrm{GE}}$ is a global efficiency indicator that
quantifies quantum Hamiltonian evolutions in terms of departures from paths of
shortest length \cite{anandan90}. Furthermore, as we have explicitly shown, it
depends in an essential manner only on the transverse magnetic field vector
$\mathbf{h}_{\perp}\left(  t\right)  $ that specifies the magnetic field
vector $\mathbf{h}\left(  t\right)  \overset{\text{def}}{=}\mathbf{h}%
_{\parallel}\left(  t\right)  +\mathbf{h}_{\perp}\left(  t\right)  $ in the
Hamiltonian \textrm{H}$\left(  t\right)  \overset{\text{def}}{=}%
\mathbf{h}\left(  t\right)  \cdot\mathbf{\boldsymbol{\sigma}}$. Second, the
speed efficiency $\eta_{\mathrm{SE}}$ is a local efficiency quantifier that
characterizes quantum evolutions in terms of deviations from paths of minimal
waste of energy \cite{uzdin12}. Furthermore, unlike $\eta_{\mathrm{GE}}$, we
have demonstrated that it depends on both the parallel and the transverse
magnetic field vectors $\mathbf{h}_{\parallel}\left(  t\right)  $ and
$\mathbf{h}_{\perp}\left(  t\right)  $, respectively, that define the magnetic
field vector $\mathbf{h}\left(  t\right)  \overset{\text{def}}{=}%
\mathbf{h}_{\parallel}\left(  t\right)  +\mathbf{h}_{\perp}\left(  t\right)  $
in the Hamiltonian \textrm{H}$\left(  t\right)  \overset{\text{def}%
}{=}\mathbf{h}\left(  t\right)  \cdot\mathbf{\boldsymbol{\sigma}}$. Third, for
stationary evolutions, unit geodesic efficiency Hamiltonian evolutions are
characterized by magnetic field vectors $\mathbf{h}\left(  t\right)  $ that
are constantly orthogonal to the Bloch vector $\mathbf{a}\left(  t\right)  $
of the evolving state vector $\left\vert \psi\left(  t\right)  \right\rangle
$. In the time-varying case, instead, the condition $\mathbf{a}\left(
t\right)  \mathbf{\cdot h}\left(  t\right)  \mathbf{=0}$ is neither necessary
(see the Fourth Example), nor sufficient (see the Second Example) to achieve
geodesicity. Fourth, in both stationary and nonstationary Hamiltonian
evolutions, nonzero trace Hamiltonians \textrm{H}$\left(  t\right)
\overset{\text{def}}{=}h_{0}(t)\mathbf{1+h}\left(  t\right)  \cdot
\mathbf{\boldsymbol{\sigma}}$ are necessarily energy wasteful. Instead,
traceless Hamiltonians \textrm{H}$\left(  t\right)  \overset{\text{def}%
}{=}\mathbf{h}\left(  t\right)  \cdot\mathbf{\boldsymbol{\sigma}}$ for which
the parallel magnetic field vector $\mathbf{h}_{\parallel}\left(  t\right)  $
of the magnetic field $\mathbf{h}\left(  t\right)  $ vanishes are not energy
wasteful. Fifth, the notion of hybrid efficiency $\eta_{\mathrm{SE}}$ offers a
new perspective on quantum evolutions, encompassing both time-optimality and
energy-optimality. In particular, it suggests clever ways to calculate the
curvature coefficient $\kappa_{\mathrm{AC}}^{2}$ of quantum evolutions
\cite{alsing24A,alsing24B} by exploiting ideas from Uzdin's speed efficiency
theoretical setting. Based on these findings, we have determined that a
practical approach to achieving quantum evolutions with unit hybrid efficiency
using time-dependent Hamiltonians involves a transverse magnetic field that
varies only in intensity, not direction. Furthermore, thinking in terms of
hybrid efficiency $\eta_{\mathrm{HE}}$ offers efficient ways to construct
speed efficient quantum evolutions starting from more speed inefficient
quantum evolutions, while preserving the same level of geodesic efficiency
$\eta_{\mathrm{GE}}$. Again, this is achieved by mixing together ideas from
Uzdin's speed efficiency theoretical setting with concepts from the
Anandan-Aharonov geometric approach to quantum evolutions as illustrated in
our work.

\medskip

We see three clear research lines that can originate from the limitations of
our investigation. A first one would be the nontrivial extension of our ideas
to multi-level quantum systems in pure states
\cite{jakob01,kimura03,krammer08,kurzy11,xie20,siewert21}. A second one would
be shifting the focus on quantum evolutions inside the Bloch sphere, where
quantum systems are in mixed quantum states. In this context, one of the
issues could be the choice of the \textquotedblleft proper\textquotedblright%
\ metric
\cite{bures69,uhlmann76,hubner92,erik20,hornedal22,cafaro23EPJ,alsing24CC,nade24}%
. A third one, the most intriguing in our view, would be studying the behavior
of the complexity of quantum evolutions
\cite{brown17,chapman18,brown19,auzzi21,cafaroPRE22,vijay22,carloPRD,aan}
(even for two-level systems) in terms of deviations from ideality specified by
curvature and efficiency concepts like the ones used in this paper. In this
context, the pertinent findings derived from some of the concepts presented in
our current work are already being successfully implemented in Refs.
\cite{carlofuture1} and \cite{carlofuture}.

\medskip

In conclusion, despite the current limitations, we are firmly convinced that
our study will motivate other scholars and pave the way for further
comprehensive investigations into the relationship between geometry and
quantum mechanics where the concept of efficiency plays a significant
role.\bigskip

\begin{acknowledgments}
The authors express their gratitude to the anonymous Referees for their
constructive feedback, which has contributed to the enhancement of the
paper\textbf{. }Any opinions, findings and conclusions or recommendations
expressed in this material are those of the author(s) and do not necessarily
reflect the views of their home Institutions.
\end{acknowledgments}

\appendix

\section{Calculation of $\Delta E$}

In this Appendix, we present a non-geometric calculation of the energy
uncertainty $\Delta E\left(  \alpha\right)  $ introduced in the first
subsection of Section III by means of ordinary quantum-mechanical rules.

Remember that the energy uncertainty $\Delta E$ is defined as%
\begin{equation}
\Delta E\overset{\text{def}}{=}\sqrt{\left\langle \mathrm{H}_{\mathrm{sub}%
\text{-\textrm{opt}}}^{2}\right\rangle -\left\langle \mathrm{H}_{\mathrm{sub}%
\text{-\textrm{opt}}}\right\rangle ^{2}}=\sqrt{\mathrm{tr}\left(
\rho\mathrm{H}_{\mathrm{sub}\text{-\textrm{opt}}}^{2}\right)  -\left[
\mathrm{tr}\left(  \rho\mathrm{H}_{\mathrm{sub}\text{-\textrm{opt}}}\right)
\right]  ^{2}}\text{.} \label{ECR}%
\end{equation}
Furthermore, recall that the sub-optimal Hamiltonian $\mathrm{H}%
_{\mathrm{sub}\text{-\textrm{opt}}}$ in Eq. (\ref{ECR}) is given by,
\begin{equation}
\mathrm{H}_{\mathrm{sub}\text{-\textrm{opt}}}\overset{\text{def}}{=}E\left[
\cos\left(  \alpha\right)  \frac{\hat{a}+\hat{b}}{2\cos\left(  \frac
{\theta_{AB}}{2}\right)  }+\sin\left(  \alpha\right)  \frac{\hat{a}\times
\hat{b}}{\sin\left(  \theta_{AB}\right)  }\right]  \cdot
\mathbf{\boldsymbol{\sigma}}\text{,}%
\end{equation}
with $\cos\left(  \theta_{AB}\right)  =\hat{a}\cdot\hat{b}$. Using standard
properties of the Pauli matrices together with the identity $\left(  \vec
{a}\cdot\mathbf{\boldsymbol{\sigma}}\right)  \left(  \vec{b}\cdot
\mathbf{\boldsymbol{\sigma}}\right)  =\left(  \vec{a}\cdot\vec{b}\right)
\mathbf{1+}i\left(  \vec{a}\times\vec{b}\right)  \cdot
\mathbf{\boldsymbol{\sigma}}$ for any pair of vectors $\vec{a}$ and $\vec{b}$
in $%
\mathbb{R}
^{3}$, we proceed with the calculation of expectation values with respect to
the initial state $\left\vert A\right\rangle $ with $\rho\overset{\text{def}%
}{=}\left\vert A\right\rangle \left\langle A\right\vert =(\mathbf{1+}\hat
{a}\cdot\mathbf{\boldsymbol{\sigma}})/2$. We note that,
\begin{align}
\left\langle \mathrm{H}_{\mathrm{sub}\text{-\textrm{opt}}}\right\rangle  &
=\mathrm{tr}\left(  \rho\mathrm{H}_{\mathrm{sub}\text{-\textrm{opt}}}\right)
\nonumber\\
&  =\mathrm{tr}\left\{  \left(  \frac{\mathbf{1+}\hat{a}\cdot
\mathbf{\boldsymbol{\sigma}}}{2}\right)  E\left[  \cos\left(  \alpha\right)
\frac{\hat{a}+\hat{b}}{2\cos\left(  \frac{\theta_{AB}}{2}\right)  }%
+\sin\left(  \alpha\right)  \frac{\hat{a}\times\hat{b}}{\sin\left(
\theta_{AB}\right)  }\right]  \cdot\mathbf{\boldsymbol{\sigma}}\right\}
\nonumber\\
&  =\mathrm{tr}\left[  \left(  \frac{\mathbf{1+}\hat{a}\cdot\vec{\sigma}}%
{2}\right)  \frac{E\cos\left(  \alpha\right)  }{2\cos\left(  \frac{\theta
_{AB}}{2}\right)  }\left(  \hat{a}+\hat{b}\right)  \cdot
\mathbf{\boldsymbol{\sigma}}\right]  +\mathrm{tr}\left[  \left(
\frac{\mathbf{1+}\hat{a}\cdot\mathbf{\boldsymbol{\sigma}}}{2}\right)
\frac{E\sin\left(  \alpha\right)  }{\sin\left(  \theta_{AB}\right)  }\left(
\hat{a}\times\hat{b}\right)  \cdot\mathbf{\boldsymbol{\sigma}}\right]
\nonumber\\
&  =\mathrm{tr}\left\{  \frac{E\cos\left(  \alpha\right)  }{4\cos\left(
\frac{\theta_{AB}}{2}\right)  }\left[  \left(  \hat{a}\cdot
\mathbf{\boldsymbol{\sigma}}\right)  \left(  \left(  \hat{a}+\hat{b}\right)
\cdot\mathbf{\boldsymbol{\sigma}}\right)  \right]  \right\} \nonumber\\
&  =\frac{E\cos\left(  \alpha\right)  }{4\cos\left(  \frac{\theta_{AB}}%
{2}\right)  }\mathrm{tr}\left\{  \hat{a}\cdot\left(  \hat{a}+\hat{b}\right)
\mathbf{1+}\hat{a}\times\left(  \hat{a}+\hat{b}\right)
\mathbf{\boldsymbol{\sigma}}\right\} \nonumber\\
&  =\frac{E\cos\left(  \alpha\right)  }{2\cos\left(  \frac{\theta_{AB}}%
{2}\right)  }\left[  \hat{a}\cdot\left(  \hat{a}+\hat{b}\right)  \right]
\text{.} \label{game}%
\end{align}
Therefore, from Eq. (\ref{game}), $\left\langle \mathrm{H}_{\mathrm{sub}%
\text{-\textrm{opt}}}\right\rangle ^{2}$ reduces to%
\begin{align}
\left\langle \mathrm{H}_{\mathrm{sub}\text{-\textrm{opt}}}\right\rangle ^{2}
&  =\left[  \frac{E\cos\left(  \alpha\right)  }{2\cos\left(  \frac{\theta
_{AB}}{2}\right)  }\right]  ^{2}\left[  \hat{a}\cdot\left(  \hat{a}+\hat
{b}\right)  \right]  ^{2}\nonumber\\
&  =\frac{E^{2}\cos^{2}\left(  \alpha\right)  }{4\cos^{2}\left(  \frac
{\theta_{AB}}{2}\right)  }\left(  \hat{a}\cdot\hat{a}+\hat{a}\cdot\hat
{b}\right)  ^{2}\nonumber\\
&  =\frac{E^{2}\cos^{2}\left(  \alpha\right)  }{4\cos^{2}\left(  \frac
{\theta_{AB}}{2}\right)  }\left[  1+\cos\left(  \theta_{AB}\right)  \right]
^{2}\nonumber\\
&  =\frac{E^{2}\cos^{2}\left(  \alpha\right)  }{4\cos^{2}\left(  \frac
{\theta_{AB}}{2}\right)  }\left[  2\cos^{2}\left(  \frac{\theta_{AB}}%
{2}\right)  \right]  ^{2}\nonumber\\
&  =E^{2}\cos^{2}\left(  \alpha\right)  \cos^{2}\left(  \frac{\theta_{AB}}%
{2}\right)  \text{,} \label{68}%
\end{align}
that is,%
\begin{equation}
\left\langle \mathrm{H}_{\mathrm{sub}\text{-\textrm{opt}}}\right\rangle
^{2}=E^{2}\cos^{2}\left(  \alpha\right)  \cos^{2}\left(  \frac{\theta_{AB}}%
{2}\right)  \text{.}%
\end{equation}
Furthermore, following a similar line of reasoning, we have
\begin{align}
\mathrm{H}_{\mathrm{sub}\text{-\textrm{opt}}}^{2}  &  =\frac{E^{2}\cos
^{2}\left(  \alpha\right)  }{4\cos\left(  \frac{\theta_{AB}}{2}\right)
}\left[  \left(  \hat{a}+\hat{b}\right)  \cdot\mathbf{\boldsymbol{\sigma}%
}\right]  \left[  \left(  \hat{a}+\hat{b}\right)  \cdot
\mathbf{\boldsymbol{\sigma}}\right]  +\nonumber\\
&  +\frac{E^{2}\sin^{2}\left(  \alpha\right)  }{\sin^{2}\left(  \theta
_{AB}\right)  }\left[  \left(  \hat{a}\times\hat{b}\right)  \cdot
\mathbf{\boldsymbol{\sigma}}\right]  \left[  \left(  \hat{a}\times\hat
{b}\right)  \cdot\mathbf{\boldsymbol{\sigma}}\right]  +\nonumber\\
&  +\frac{E^{2}\sin\left(  \alpha\right)  \cos\left(  \alpha\right)  }%
{2\sin\left(  \theta_{AB}\right)  \cos\left(  \frac{\theta_{AB}}{2}\right)
}\left\{  \left[  \left(  \hat{a}+\hat{b}\right)  \cdot
\mathbf{\boldsymbol{\sigma}}\right]  \left[  \left(  \hat{a}\times\hat
{b}\right)  \cdot\mathbf{\boldsymbol{\sigma}}\right]  +\left[  \left(  \hat
{a}\times\hat{b}\right)  \cdot\mathbf{\boldsymbol{\sigma}}\right]  \left[
\left(  \hat{a}+\hat{b}\right)  \cdot\mathbf{\boldsymbol{\sigma}}\right]
\right\} \nonumber\\
&  =\frac{E^{2}\cos^{2}\left(  \alpha\right)  }{4\cos\left(  \frac{\theta
_{AB}}{2}\right)  }\left(  \hat{a}+\hat{b}\right)  ^{2}\mathbf{1+}\frac
{E^{2}\sin^{2}\left(  \alpha\right)  }{\sin^{2}\left(  \theta_{AB}\right)
}\left(  \hat{a}\times\hat{b}\right)  ^{2}\mathbf{1+}\frac{E^{2}\sin\left(
\alpha\right)  \cos\left(  \alpha\right)  }{2\sin\left(  \theta_{AB}\right)
\cos\left(  \frac{\theta_{AB}}{2}\right)  }\cdot\nonumber\\
&  \cdot\left[
\begin{array}
[c]{c}%
\left(  \hat{a}+\hat{b}\right)  \cdot\left(  \hat{a}\times\hat{b}\right)
\mathbf{1+}i\left(  \left(  \hat{a}+\hat{b}\right)  \times\left(  \hat
{a}\times\hat{b}\right)  \right)  \cdot\mathbf{\boldsymbol{\sigma}}+\\
+\left(  \hat{a}\times\hat{b}\right)  \cdot\left(  \hat{a}+\hat{b}\right)
\mathbf{1+}i\left(  \left(  \hat{a}\times\hat{b}\right)  \times\left(  \hat
{a}+\hat{b}\right)  \right)  \cdot\mathbf{\boldsymbol{\sigma}}%
\end{array}
\right] \nonumber\\
&  =\frac{E^{2}\cos^{2}\left(  \alpha\right)  }{4\cos\left(  \frac{\theta
_{AB}}{2}\right)  }\left(  \hat{a}+\hat{b}\right)  ^{2}\mathbf{1+}\frac
{E^{2}\sin^{2}\left(  \alpha\right)  }{\sin^{2}\left(  \theta_{AB}\right)
}\left(  \hat{a}\times\hat{b}\right)  ^{2}\mathbf{1}\nonumber\\
&  =\frac{E^{2}\cos^{2}\left(  \alpha\right)  }{4\cos\left(  \frac{\theta
_{AB}}{2}\right)  }4\cos\left(  \frac{\theta_{AB}}{2}\right)  \mathbf{1+}%
\frac{E^{2}\sin^{2}\left(  \alpha\right)  }{\sin^{2}\left(  \theta
_{AB}\right)  }\sin^{2}\left(  \theta_{AB}\right)  \mathbf{1}\nonumber\\
&  =E^{2}\mathbf{1}\text{,}%
\end{align}
therefore, $\left\langle \mathrm{H}_{\mathrm{sub}\text{-\textrm{opt}}}%
^{2}\right\rangle $ becomes%
\begin{align}
\left\langle \mathrm{H}_{\mathrm{sub}\text{-\textrm{opt}}}^{2}\right\rangle
&  =\mathrm{tr}\left[  \left(  \frac{\mathbf{1+}\hat{a}\cdot
\mathbf{\boldsymbol{\sigma}}}{2}\right)  E^{2}\mathbf{1}\right] \nonumber\\
&  =\mathrm{tr}\left(  \frac{E^{2}}{2}\mathbf{1}\right) \nonumber\\
&  =E^{2}\text{.} \label{68B}%
\end{align}
In conclusion, substituting Eqs. (\ref{68}) and (\ref{68B}) into Eq.
(\ref{ECR}), the energy uncertainty $\Delta E$ becomes%
\begin{equation}
\Delta E=\Delta E\left(  \alpha\right)  \overset{\text{def}}{=}E\sqrt
{1-\cos^{2}\left(  \alpha\right)  \cos^{2}\left(  \frac{\theta_{AB}}%
{2}\right)  }\text{.} \label{bruno}%
\end{equation}
The derivation of Eq. (\ref{bruno}) ends our discussion here.

\section{Calculation of $\kappa_{\mathrm{AC}}^{2}$}

In this Appendix, we introduce the essential elements yielding the notion of
curvature coefficient $\kappa_{\mathrm{AC}}^{2}$ of a quantum evolution
\cite{alsing24A,alsing24B}. In particular, this discussion leads to the
expression for $\kappa_{\mathrm{AC}}^{2}$ in Eq. (\ref{xxx}).

Assume to consider a time-varying Hamiltonian evolution specified by the
Schr\"{o}dinger equation $i\hslash\partial_{t}\left\vert \psi\left(  t\right)
\right\rangle =\mathrm{H}\left(  t\right)  \left\vert \psi\left(  t\right)
\right\rangle $, where the normalized quantum state $\left\vert \psi\left(
t\right)  \right\rangle $ belongs to an arbitrary $N$-dimensional complex
Hilbert space $\mathcal{H}_{N}$. Usually, $\left\vert \psi\left(  t\right)
\right\rangle $ satisfies $\left\langle \psi\left(  t\right)  \left\vert
\dot{\psi}\left(  t\right)  \right.  \right\rangle =(-i/\hslash)\left\langle
\psi\left(  t\right)  \left\vert \mathrm{H}\left(  t\right)  \right\vert
\psi\left(  t\right)  \right\rangle \neq0$. From the state $\left\vert
\psi\left(  t\right)  \right\rangle $, we construct the parallel transported
unit state vector $\left\vert \Psi\left(  t\right)  \right\rangle
\overset{\text{def}}{=}e^{i\beta\left(  t\right)  }\left\vert \psi\left(
t\right)  \right\rangle $ with the phase $\beta\left(  t\right)  $ being such
that $\left\langle \Psi\left(  t\right)  \left\vert \dot{\Psi}\left(
t\right)  \right.  \right\rangle =0$. Notice that $i\hslash\left\vert
\dot{\Psi}\left(  t\right)  \right\rangle =\left[  \mathrm{H}\left(  t\right)
-\hslash\dot{\beta}\left(  t\right)  \right]  \left\vert \Psi\left(  t\right)
\right\rangle $. Then, the condition $\left\langle \Psi\left(  t\right)
\left\vert \dot{\Psi}\left(  t\right)  \right.  \right\rangle =0$ is identical
to having $\beta\left(  t\right)  $ equal to%
\begin{equation}
\beta\left(  t\right)  \overset{\text{def}}{=}\frac{1}{\hslash}\int_{0}%
^{t}\left\langle \psi\left(  t^{\prime}\right)  \left\vert \mathrm{H}\left(
t^{\prime}\right)  \right\vert \psi\left(  t^{\prime}\right)  \right\rangle
dt^{\prime}\text{.}%
\end{equation}
Therefore, the state vector $\left\vert \Psi\left(  t\right)  \right\rangle $
becomes%
\begin{equation}
\left\vert \Psi\left(  t\right)  \right\rangle =e^{(i/\hslash)\int_{0}%
^{t}\left\langle \psi\left(  t^{\prime}\right)  \left\vert \mathrm{H}\left(
t^{\prime}\right)  \right\vert \psi\left(  t^{\prime}\right)  \right\rangle
dt^{\prime}}\left\vert \psi\left(  t\right)  \right\rangle \text{,}%
\end{equation}
and fulfills the evolution equation $i\hslash\left\vert \dot{\Psi}\left(
t\right)  \right\rangle =\Delta\mathrm{H}\left(  t\right)  \left\vert
\Psi\left(  t\right)  \right\rangle $ with $\Delta\mathrm{H}\left(  t\right)
\overset{\text{def}}{=}\mathrm{H}\left(  t\right)  -\left\langle
\mathrm{H}\left(  t\right)  \right\rangle $. As mentioned in Ref.
\cite{alsing24B}, the speed $v(t)$ of a quantum evolution is not constant when
the Hamiltonian changes in time. Specifically, $v(t)$ satisfies $v^{2}\left(
t\right)  =\left\langle \dot{\Psi}\left(  t\right)  \left\vert \dot{\Psi
}\left(  t\right)  \right.  \right\rangle =\left\langle \left(  \Delta
\mathrm{H}\left(  t\right)  \right)  ^{2}\right\rangle /\hslash^{2}$. For
convenience, we present the arc length $s=s\left(  t\right)  $ defined in
terms of $v\left(  t\right)  $ as%
\begin{equation}
s\left(  t\right)  \overset{\text{def}}{=}\int_{0}^{t}v(t^{\prime})dt^{\prime
}\text{,} \label{s-equation}%
\end{equation}
where $ds=v(t)dt$, that is, $\partial_{t}=v(t)\partial_{s}$. Lastly,
introducing the adimensional operator $\Delta h\left(  t\right)
\overset{\text{def}}{=}\Delta\mathrm{H}\left(  t\right)  /[\hslash
v(t)]=\Delta\mathrm{H}\left(  t\right)  /\sqrt{\left\langle \left(
\Delta\mathrm{H}\left(  t\right)  \right)  ^{2}\right\rangle }$, the
normalized tangent vector $\left\vert T\left(  s\right)  \right\rangle
\overset{\text{def}}{=}\partial_{s}\left\vert \Psi\left(  s\right)
\right\rangle =\left\vert \Psi^{\prime}\left(  s\right)  \right\rangle $
reduces to $\left\vert T\left(  s\right)  \right\rangle =-i\Delta h\left(
s\right)  \left\vert \Psi\left(  s\right)  \right\rangle $. We point out that
$\left\langle T\left(  s\right)  \left\vert T\left(  s\right)  \right.
\right\rangle =1$ by construction and, additionally, $\partial_{s}\left\langle
\Delta h(s)\right\rangle =\left\langle \Delta h^{\prime}(s)\right\rangle $.
Interestingly, this latter relation remains valid for arbitrary powers of
differentiation. For instance, to the second power, we find $\partial_{s}%
^{2}\left\langle \Delta h(s)\right\rangle =\left\langle \Delta h^{\prime
\prime}(s)\right\rangle $. We can generate $\left\vert T^{\prime}\left(
s\right)  \right\rangle \overset{\text{def}}{=}\partial_{s}\left\vert T\left(
s\right)  \right\rangle $ from the tangent vector $\left\vert T\left(
s\right)  \right\rangle =-i\Delta h\left(  s\right)  \left\vert \Psi\left(
s\right)  \right\rangle $. More explicitly, we get $\left\vert T^{\prime
}\left(  s\right)  \right\rangle =-i\Delta h(s)\left\vert \Psi^{\prime}\left(
s\right)  \right\rangle -i\Delta h^{\prime}(s)\left\vert \Psi\left(  s\right)
\right\rangle $ where, in general,
\begin{equation}
\left\langle T^{\prime}\left(  s\right)  \left\vert T^{\prime}\left(
s\right)  \right.  \right\rangle =\left\langle \left(  \Delta h^{\prime
}(s)\right)  ^{2}\right\rangle +\left\langle \left(  \Delta h(s)\right)
^{4}\right\rangle -2i\operatorname{Re}\left[  \left\langle \Delta h^{\prime
}(s)\left(  \Delta h(s)\right)  ^{2}\right\rangle \right]  \neq1\text{.}%
\end{equation}
We are now ready to introduce the curvature coefficient for quantum evolutions
emerging from time-varying Hamiltonians. As a matter of fact, having presented
the vectors $\left\vert \Psi\left(  s\right)  \right\rangle $, $\left\vert
T\left(  s\right)  \right\rangle $, and $\left\vert T^{\prime}\left(
s\right)  \right\rangle $, we can finally define the curvature coefficient
$\kappa_{\mathrm{AC}}^{2}\left(  s\right)  \overset{\text{def}}{=}\left\langle
\tilde{N}_{\ast}\left(  s\right)  \left\vert \tilde{N}_{\ast}\left(  s\right)
\right.  \right\rangle $ with $\left\vert \tilde{N}_{\ast}\left(  s\right)
\right\rangle \overset{\text{def}}{=}\mathrm{P}^{\left(  \Psi\right)
}\left\vert T^{\prime}\left(  s\right)  \right\rangle $, $\mathrm{P}^{\left(
\Psi\right)  }\overset{\text{def}}{=}\mathrm{I}-\left\vert \Psi\left(
s\right)  \right\rangle \left\langle \Psi\left(  s\right)  \right\vert $, and
\textquotedblleft$\mathrm{I}$\textquotedblright\ being the identity operator
in $\mathcal{H}_{N}$. As reported in Refs. \cite{alsing24A,alsing24B}, the
subscript \textquotedblleft\textrm{AC}\textquotedblright\textrm{\ }stands for
Alsing and Cafaro. Note that the curvature coefficient $\kappa_{\mathrm{AC}%
}^{2}\left(  s\right)  $ can be rewritten as%
\begin{equation}
\kappa_{\mathrm{AC}}^{2}\left(  s\right)  \overset{\text{def}}{=}\left\Vert
\mathrm{D}\left\vert T(s)\right\rangle \right\Vert ^{2}=\left\Vert
\mathrm{D}^{2}\left\vert \Psi\left(  s\right)  \right\rangle \right\Vert
^{2}\text{,} \label{peggio}%
\end{equation}
with $\mathrm{D}\overset{\text{def}}{=}\mathrm{P}^{\left(  \Psi\right)
}d/ds=\left(  \mathrm{I}-\left\vert \Psi\right\rangle \left\langle
\Psi\right\vert \right)  d/ds$ and $\mathrm{D}\left\vert T(s)\right\rangle
\overset{\text{def}}{=}\mathrm{P}^{\left(  \Psi\right)  }\left\vert T^{\prime
}(s)\right\rangle $ denoting the covariant derivative
\cite{cafaro23,samuel88,paulPRA23}. We observe that the curvature coefficient
$\kappa_{\mathrm{AC}}^{2}\left(  s\right)  $ in Eq. (\ref{peggio}) equals the
magnitude squared of the second covariant derivative of the state vector
$\left\vert \Psi\left(  s\right)  \right\rangle $ used to construct the
quantum Schr\"{o}dinger trajectory in projective Hilbert space. To be crystal
clear, we remark that $\left\vert \tilde{N}_{\ast}\left(  s\right)
\right\rangle $ is a vector that is neither orthogonal to the vector
$\left\vert T\left(  s\right)  \right\rangle $ nor normalized to one. On the
contrary, despite being unnormalized, $\left\vert \tilde{N}\left(  s\right)
\right\rangle \overset{\text{def}}{=}\mathrm{P}^{\left(  T\right)  }%
\mathrm{P}^{\left(  \Psi\right)  }\left\vert T^{\prime}\left(  s\right)
\right\rangle $ is orthogonal to $\left\vert T\left(  s\right)  \right\rangle
$. Lastly, $\left\vert N\left(  s\right)  \right\rangle \overset{\text{def}%
}{=}$ $\left\vert \tilde{N}\left(  s\right)  \right\rangle /\sqrt{\left\langle
\tilde{N}\left(  s\right)  \left\vert \tilde{N}\left(  s\right)  \right.
\right\rangle }$ is a normalized vector which is orthogonal to $\left\vert
T\left(  s\right)  \right\rangle $. Summing up, $\left\{  \left\vert
\Psi\left(  s\right)  \right\rangle \text{, }\left\vert T\left(  s\right)
\right\rangle \text{, }\left\vert N\left(  s\right)  \right\rangle \right\}  $
is the set of three orthonormal vectors needed to specify the curvature
coefficient of a quantum evolution. Note that, although $\mathcal{H}_{N}$ can
possess arbitrary dimension as a complex space, we limit our attention to the
three-dimensional complex subspace generated by $\left\{  \left\vert
\Psi\left(  s\right)  \right\rangle \text{, }\left\vert T\left(  s\right)
\right\rangle \text{, }\left\vert N\left(  s\right)  \right\rangle \right\}
$. Nevertheless, our working assumption agrees with the classical geometric
viewpoint according to which the curvature and torsion coefficients can be
regarded as the lowest and second-lowest, respectively, members of a family of
generalized curvatures functions \cite{alvarez19}. Particularly, for curves in
higher-dimensional spaces, this sound geometric perspective demands a set of
$m$ orthonormal vectors to construct $\left(  m-1\right)  $-generalized
curvature functions \cite{alvarez19}.

In general, the direct evaluation of the time-dependent curvature coefficient
$\kappa_{\mathrm{AC}}^{2}\left(  s\right)  $ in Eqs. (\ref{peggio}) by means
of the projection operators formalism is challenging. The difficulty is caused
by the fact that, in analogy to what happens in the classical case of space
curves in $%
\mathbb{R}
^{3}$ \cite{parker77}, there exist two main drawbacks when reparameterizing a
quantum curve by its arc length $s$. Firstly, we may be unable to evaluate in
closed form $s\left(  t\right)  $ in Eq. (\ref{s-equation}). Secondly, even if
we are capable of getting $s=s\left(  t\right)  $, we may not be able to
invert this relation and, thus, arrive at $t=t\left(  s\right)  $ needed to
obtain $\left\vert \Psi\left(  s\right)  \right\rangle \overset{\text{def}%
}{=}\left\vert \Psi\left(  t(s)\right)  \right\rangle $. To avoid these
challenges, we can rewrite $\kappa_{\mathrm{AC}}^{2}\left(  s\right)  $ in Eq.
(\ref{peggio}) by means of expectation values defined with respect to the
state $\left\vert \Psi\left(  t\right)  \right\rangle $ (or, alternatively,
with respect to $\left\vert \psi\left(  t\right)  \right\rangle $) that can be
computed without the relation $t=t\left(  s\right)  $. For ease of notation,
in the following discussion, we shall avoid making any explicit reference to
the the $s$-dependence of the variety of operators and expectation values
being considered. For instance, $\Delta h\left(  s\right)  $ will appear as
$\Delta h$. After some algebra, we get%
\begin{equation}
\left\vert \tilde{N}_{\ast}\right\rangle =-\left\{  \left[  \left(  \Delta
h\right)  ^{2}-\left\langle \left(  \Delta h\right)  ^{2}\right\rangle
\right]  +i\left[  \Delta h^{\prime}-\left\langle \Delta h^{\prime
}\right\rangle \right]  \right\}  \left\vert \Psi\right\rangle \text{,}
\label{nas}%
\end{equation}
with $\Delta h^{\prime}=\partial_{s}\left(  \Delta h\right)  =\left[
\partial_{t}\left(  \Delta h\right)  \right]  /v\left(  t\right)  $. To
calculate $\kappa_{\mathrm{AC}}^{2}\left(  s\right)  \overset{\text{def}%
}{=}\left\langle \tilde{N}_{\ast}\left(  s\right)  \left\vert \tilde{N}_{\ast
}\left(  s\right)  \right.  \right\rangle $, it is useful to employ the
Hermitian operator $\hat{\alpha}_{1}\overset{\text{def}}{=}\left(  \Delta
h\right)  ^{2}-\left\langle \left(  \Delta h\right)  ^{2}\right\rangle $ and
the anti-Hermitian operator $\hat{\beta}_{1}\overset{\text{def}}{=}i\left[
\Delta h^{\prime}-\left\langle \Delta h^{\prime}\right\rangle \right]  $ where
$\hat{\beta}_{1}^{\dagger}=-\hat{\beta}_{1}$. Then, $\left\vert \tilde
{N}_{\ast}\right\rangle =-\left(  \hat{\alpha}_{1}+\hat{\beta}_{1}\right)
\left\vert \Psi\right\rangle $ and $\left\langle \tilde{N}_{\ast}\left(
s\right)  \left\vert \tilde{N}_{\ast}\left(  s\right)  \right.  \right\rangle
$ equals $\left\langle \hat{\alpha}_{1}^{2}\right\rangle -\left\langle
\hat{\beta}_{1}^{2}\right\rangle +\left\langle \left[  \hat{\alpha}_{1}\text{,
}\hat{\beta}_{1}\right]  \right\rangle $ with $\left[  \hat{\alpha}_{1}\text{,
}\hat{\beta}_{1}\right]  \overset{\text{def}}{=}\hat{\alpha}_{1}\hat{\beta
}_{1}-\hat{\beta}_{1}\hat{\alpha}_{1}$ denoting the quantum commutator of
$\hat{\alpha}_{1}$ and $\hat{\beta}_{1}$. Note that the expectation value
$\left\langle \left[  \hat{\alpha}_{1}\text{, }\hat{\beta}_{1}\right]
\right\rangle $ belongs to $%
\mathbb{R}
$ given that $\left[  \hat{\alpha}_{1}\text{, }\hat{\beta}_{1}\right]  $ is a
Hermitian operator. This is due to the fact that $\hat{\alpha}_{1}$ and
$\hat{\beta}_{1}$ represent Hermitian and anti-Hermitian operators,
respectively. Exploiting the definitions of $\hat{\alpha}_{1}$ and $\hat
{\beta}_{1}$, we obtain $\left\langle \hat{\alpha}_{1}^{2}\right\rangle
=\left\langle (\Delta h)^{4}\right\rangle -\left\langle (\Delta h)^{2}%
\right\rangle ^{2}$, $\left\langle \hat{\beta}_{1}^{2}\right\rangle =-\left[
\left\langle (\Delta h^{\prime})^{2}\right\rangle -\left\langle \Delta
h^{\prime}\right\rangle ^{2}\right]  $, and $\left\langle \left[  \hat{\alpha
}_{1}\text{, }\hat{\beta}_{1}\right]  \right\rangle =i\left\langle \left[
(\Delta h)^{2}\text{, }\Delta h^{\prime}\right]  \right\rangle $. Observe that
$\left\langle \left[  (\Delta h)^{2}\text{, }\Delta h^{\prime}\right]
\right\rangle $ is purely imaginary given that $\left[  (\Delta h)^{2}\text{,
}\Delta h^{\prime}\right]  $ is a anti-Hermitian operator. For completeness,
we remark that $\left[  (\Delta h)^{2}\text{, }\Delta h^{\prime}\right]  $ is
generally not a null operator. As a matter of fact, $\left[  (\Delta
h)^{2}\text{, }\Delta h^{\prime}\right]  =\Delta h\left[  \Delta h\text{,
}\Delta h^{\prime}\right]  +\left[  \Delta h\text{, }\Delta h^{\prime}\right]
\Delta h$ with $\left[  \Delta h\text{, }\Delta h^{\prime}\right]  =\left[
\mathrm{H}\text{, }\mathrm{H}^{\prime}\right]  $. Then, focusing on
time-varying qubit Hamiltonians of the form \textrm{H}$\left(  s\right)
\overset{\text{def}}{=}\mathbf{h}\left(  s\right)  \cdot
\mathbf{\boldsymbol{\sigma}}$, the commutator $\left[  \mathrm{H}\text{,
}\mathrm{H}^{\prime}\right]  =2i(\mathbf{h}\times\mathbf{h}^{\prime}%
)\cdot\mathbf{\boldsymbol{\sigma}}$ is not equal to zero given that the
vectors $\mathbf{h}$ and $\mathbf{h}^{\prime}$ are not generally collinear.
Obviously, $\mathbf{\boldsymbol{\sigma}}$ is the vector operator whose
components are defined by the Pauli operators $\sigma_{x}$, $\sigma_{y}$, and
$\sigma_{z}$. Finally, a computationally useful expression for the curvature
coefficient $\kappa_{\mathrm{AC}}^{2}\left(  s\right)  $ in Eq. (\ref{peggio})
in an arbitrary time-varying framework becomes%
\begin{equation}
\kappa_{\mathrm{AC}}^{2}\left(  s\right)  =\left\langle (\Delta h)^{4}%
\right\rangle -\left\langle (\Delta h)^{2}\right\rangle ^{2}+\left[
\left\langle (\Delta h^{\prime})^{2}\right\rangle -\left\langle \Delta
h^{\prime}\right\rangle ^{2}\right]  +i\left\langle \left[  (\Delta
h)^{2}\text{, }\Delta h^{\prime}\right]  \right\rangle \text{.}
\label{curvatime}%
\end{equation}
From Eq. (\ref{curvatime}), we realize that when the Hamiltonian \textrm{H}
does not depend on time, $\Delta h^{\prime}$ reduces to the null operator and
we recuperate the stationary limit $\left\langle (\Delta h)^{4}\right\rangle
-\left\langle (\Delta h)^{2}\right\rangle ^{2}$ of $\kappa_{\mathrm{AC}}%
^{2}\left(  s\right)  $ \cite{alsing24A}.

The formulation of $\kappa_{\mathrm{AC}}^{2}\left(  s\right)  $ in Eq.
(\ref{curvatime}) is acquired by means of an approach that relies upon the
calculation of expectation values which, in turn, demand the knowledge of
\ the state vector $\left\vert \psi\left(  t\right)  \right\rangle $
satisfying the time-dependent Schr\"{o}dinger evolution equation. As remarked
in Ref. \cite{alsing24B}, such expectation-values approach yields an
insightful statistical interpretation for $\kappa_{\mathrm{AC}}^{2}\left(
s\right)  $. Simultaneously, however, this approach does not posses a neat
geometric significance. Driven by this shortfall and limiting our attention to
time-varying Hamiltonians and two-level quantum systems, it is possible to
obtain a closed-form expression for the curvature coefficient for a curve
traced out by a single-qubit quantum state that changes under the action of a
general time-dependent Hamiltonian. The curvature coefficient $\kappa
_{\mathrm{AC}}^{2}$ can be fully expressed in terms of only two real
three-dimensional vectors with a transparent geometrical interpretation.
Namely, the two vectors are the Bloch vector $\mathbf{a}\left(  t\right)  $
and the magnetic field vector $\mathbf{h}\left(  t\right)  $. While the former
vector defines the density operator $\rho\left(  t\right)  =$ $\left\vert
\psi\left(  t\right)  \right\rangle \left\langle \psi\left(  t\right)
\right\vert \overset{\text{def}}{=}\left[  \mathrm{I}+\mathbf{a}\left(
t\right)  \cdot\mathbf{\boldsymbol{\sigma}}\right]  /2$, the latter
characterizes the time-varying (traceless) Hamiltonian \textrm{H}$\left(
t\right)  \overset{\text{def}}{=}\mathbf{h}\left(  t\right)  \cdot
\mathbf{\boldsymbol{\sigma}}$. Following the extended analysis presented in
Ref. \cite{alsing24B}, we obtain%
\begin{equation}
\kappa_{\mathrm{AC}}^{2}\left(  \mathbf{a}\text{, }\mathbf{h}\right)
=4\frac{\left(  \mathbf{a\cdot h}\right)  ^{2}}{\mathbf{h}^{2}-\left(
\mathbf{a\cdot h}\right)  ^{2}}+\frac{\left[  \mathbf{h}^{2}\mathbf{\dot{h}%
}^{2}-\left(  \mathbf{h\cdot\dot{h}}\right)  ^{2}\right]  -\left[  \left(
\mathbf{a\cdot\dot{h}}\right)  \mathbf{h-}\left(  \mathbf{a\cdot h}\right)
\mathbf{\dot{h}}\right]  ^{2}}{\left[  \mathbf{h}^{2}-\left(  \mathbf{a\cdot
h}\right)  ^{2}\right]  ^{3}}+4\frac{\left(  \mathbf{a\cdot h}\right)  \left[
\mathbf{a\cdot}\left(  \mathbf{h\times\dot{h}}\right)  \right]  }{\left[
\mathbf{h}^{2}-\left(  \mathbf{a\cdot h}\right)  ^{2}\right]  ^{2}}\text{.}
\label{XXX}%
\end{equation}
The derivation of $\kappa_{\mathrm{AC}}^{2}$ in Eq. (\ref{XXX}) serves as a
valuable tool for computational analysis in qubit systems. Furthermore, it
presents a lucid geometric interpretation of the curvature of a quantum
evolution by means of the (normalized unitless) Bloch vector $\mathbf{a}$ and
the\ (usually unnormalized, with $\left[  \mathbf{h}\right]  _{\mathrm{MKSA}%
}=$\textrm{joules}$=\sec.^{-1}$ when putting $\hslash=1$) magnetic field
vector $\mathbf{h}$. Our discussion concludes with the derivation of Eq.
(\ref{XXX}).

\bigskip

\end{document}